\documentclass[journal]{IEEEtran}
\usepackage[printonlyused]{acronym}

\usepackage{subcaption}

\ifCLASSINFOpdf
  \usepackage[pdftex]{graphicx}
\else
  \usepackage[dvips]{graphicx}
\fi
\usepackage{epstopdf}

\acrodef{GNSS} {Global Navigation Satellite Systems}
\acrodef{IC} {Integrated Circuits}
\acrodef{WLAN} {Wireless Local Area Network}
\acrodef{IPR}{Intellectual Property Rights}
\acrodef{SPN} {Sensor Pattern Noise}
\acrodef{ANN} {Artificial Neural Networks}
\acrodef{LTE} {Long Term Evolution}
\acrodef {MDA} {Multiple Discriminant Analysis}
\acrodef {ML} {Maximum Likelihood}
\acrodef {FLD} {Fisher’s linear discriminant}
\acrodef {GSM} {Global System for Mobile Communications}
\acrodef {UMTS} {Universal Mobile Telecommunications System}
\acrodef {PUF} {Physical Unclonable Functions}
\acrodef{MEMS} {Micro-Electro-MEchanical Systems}
\acrodef{IOT} {Internet of Things}
\acrodef{ITA} {Inter Arrival Time}
\acrodef{NFC} {Near Field Communication}
\acrodef{IC} {Integrated Circuits}
\acrodef{SPN} {Sensor Pattern Noise}
\acrodef{RMS} {Root Mean Square}
\acrodef{RBF} {Radial Basis Function}
\acrodef{KNN} {K-Nearest Neighbor}
\acrodef{VT} 	{Variance Trajectory}
\acrodef{QP} 	{Quadratic Programming}
\acrodef{IM} 	{Interpolation Matrix}
\acrodef{OAO} {One-Against-One}
\acrodef{OAA} {One-Against-All}
\acrodef{DAGSVM} {Directed Acyclic Graph-Support Vector Machine}
\acrodef{PCA} {Principal Component Analysis}
\acrodef{HHT} {Hilbert Huang Transform}
\acrodef{ENF} {Electric Network Frequency}
\acrodef{PSD} {Power Spectrum Density}
\acrodef{AP} {Access Point}
\acrodef{DRA} {Dimensional Reduction Analysis}
\acrodef{MDA} {Multiple discriminant analysis (MDA)}
\acrodef{ML} {maximum likelihood (ML)}
\acrodef{CFA} {Color Filter Array} 
\acrodef{GLCM} {Gray level Co-occurrence Matrix}
\acrodef{CMOS} {complementary metal\-oxide semiconductor}
\acrodef{MAC} {Medium Access Control}
\acrodef{CCD} {Charge Coupled Device}
\acrodef{ENF} {Electric Network Frequency}
\acrodef{WiMAX} {Wireless Medium Access}
\acrodef{RF} {Radio Frequency}
\acrodef{ISO} {International Organization for Standardization}
\acrodef{CMOS} {Complementary Metal Oxide Semiconductor}
\acrodef{JPEG} {Joint Photographic Experts Group}
\acrodef{PRNU} {photo-response non-uniformity noise}
\acrodef{SNR} {Signal Noise Ratio}
\acrodef{PNU} {Pixel Non-Uniformity}
\acrodef{SVM} {Support Vector Machine}
\acrodef{FLD} {Fishers linear discriminant}
\acrodef{MFCC} {Mel-Frequency Cepstrum Coefficient}
\acrodef{GRLVQI} {Generalized Relevance Learning Vector Quantization Improved} 
\acrodef{WLPI} {Wireless Physical Layer Identification}
\acrodef{GLDS} {Generalized linear discriminant sequence}
\acrodef{VQ} {Vector Quantizer}
\acrodef{GPS} {Global Positioning Systems}
\acrodef{LPCC} {Linear Predictive Cepstrum Coefficient}
\acrodef{PLPC} {Perceptually-based Linear Predictive Coefficients}
\acrodef{SRC} {Sparse representations Classifier}
\acrodef{NN} {Nearest Neighbour}
\acrodef{VP} {Virtual Proofs}
\acrodef{PCA} {Principal Component Analysis}
\acrodef{UBM} {Universal Background Mode}
\acrodef{GMM} {Gaussian mixture model}
\acrodef{PCM} {Pulse Code Modulation}
\acrodef{MP3} {MPEG-2 Audio Layer 3}
\acrodef{USRP} {Universal Software Radio Peripheral}
\acrodef{AAC} {Advanced Audio Coding}
\acrodef{RFID} {Radio Frequency IDentifier}
\acrodef{TCP} {Transport Communication Protocol}
\acrodef{UDP} {User Datagram Protocol}
\acrodef{IoT} {Internet of Things}
\acrodef{ADC} {Analog Digital Converter}
\acrodef{RBF} {Radial Basis Function}
\acrodef{GLONASS} {GLObal NAvigation Satellite System}
\acrodef{LCD} {Liquid-Crystal Display}
\acrodef{WARP} {Wireless Open-Access Research Platform}
\acrodef{SDR} {Software Defined Radio}
\acrodef{EVM} {Error Vector Magnitude}
\acrodef{FFT} {Fast Fourier Transform}
\acrodef{SFS} {Sequential Feature Selection}
\acrodef{ROC} {Receiver Operating Characteristics}
\acrodef{EER} {Equal Error Rate}
\acrodef{OAO} {One Against One}
\acrodef{OAA} {One Against All}
\acrodef{SMO} {Sequential minimal optimization}
\acrodef{AWGN} {Additive white Gaussian noise}

\usepackage{subfig}
\usepackage{adjustbox}
\usepackage{array}
\usepackage{rotating}

\newcolumntype{R}[2]{%
    >{\adjustbox{angle=#1,lap=\width-(#2)}\bgroup}%
    l%
    <{\egroup}%
}

\begin{document}

%
\title{Mobile phone identification through the built-in magnetometers}

\author{
  Gianmarco~Baldini$^{*}$, Gary~Steri$^{*}$, Raimondo~Giuliani$^{*}$,Vladimir~Kyovtorov$^{*}$\\ 
$^{*}$European Commission, Joint Research Centre, Ispra, Italy\\
gianmarco.baldini@jrc.ec.europa.eu,  gary.steri@jrc.ec.europa.eu, raimondo.giuliani@jrc.ec.europa.eu, vladimir.kyovtorov@jrc.ec.europa.eu}

\maketitle

\begin{abstract}
Mobile phones identification through their built-in components has been demonstrated in literature for various types of sensors including the camera, microphones and accelerometers. The identification is performed by the exploitation of the small but significant differences in the electronic circuits generated during the production process. Thus, these differences become an intrinsic property of the electronic components, which can be detected and become an unique fingerprint of the component and of the mobile phone. In this paper, we investigate the identification of mobile phones through their built-in magnetometers, which has not been reported in literature yet. Magnetometers are stimulated with different waveforms using a solenoid connected to a computer's audio board. The identification is performed analyzing the digital output of the magnetometer through the use of statistical features and the Support Vector Machine (SVM) machine learning algorithm. We prove that this technique can distinguish different models and brands with very high accuracy but it can only distinguish phones of the same model with limited accuracy.
\end{abstract}

\section{Introduction}
The mobile phone identification based on the physical properties of its components is related to the capability to distinguish between phones of the same model but different serial numbers (intra-model identification) and between phones of different models and brands (inter-model identification). Intra-model identification is usually more difficult to achieve than inter-model identification because mobile phone manufacturers use the same materials in the same model, while different models are usually built using different materials and components.

The identification can be performed by exploiting the tiny but significant differences in the materials or the differences introduced in the manufacturing process. These differences result in small disturbances in the digital output generated by the mobile phone. They are also called \textit{fingerprints} of the component, similarly to the fingerprints of a human being. For example, the digital camera of a mobile phone introduce in every image processed and stored in the memory of the phone, a specific feature which is called \ac{SPN}. The \ac{SPN} can be extracted from a sufficient number of images as demonstrated in the pioneering work by \cite{lukasspn}.
In a similar way, researchers have demonstrated the capability to identify mobile phones with different degrees of accuracy for the built-in accelerometers, radio frequency components, microphones and so on. The section \ref{relatedwork} gives an overview of the research work for the different types of components and the different adopted techniques. 

The identification of mobile phones through their components has various applications in security, forensics or fight against the counterfeiting of electronic products. 
In security, identification proofs based on physical characteristics are much more difficult to be faked and reproduced and they are intrinsically related to the component and the mobile phone itself. Fingerprints can be used to perform multi-factor authentication where physical identification is combined with cryptographic authentication (see \cite{suski2008using}). In this case, both inter-model and intra-model identification would be applicable.  
Many forensics applications can exploit the knowledge of the identity of the mobile phone used in a crime scene. For example, camera identification techniques based on \ac{SPN} can be applied to the images uploaded to the web by a criminal to identify the mobile phone and the criminal identity \cite{Fridrich}.
In the fight against the distribution of counterfeit products, the identification of a mobile phone through its components can be used to detect counterfeit phones, which are often built with cheaper components compared to the original ones. Cheaper components would have a different and distinct signatures from the ones used in original phones. In this case, inter-model identification would be appropriate.
In fingerprinting literature, a distinction is made between \textit{Classification} and \textit{Verification}. In this paper, we will adopt similar definitions to what presented in \cite{Bihl} where Verification is the process to verify the claimed identity of a mobile phone, while Classification considers authorized mobile phone fingerprints (e.g., a reference library of known mobile phones) to create a model that best discriminates the mobile phones on the basis of the built-in magnetometers. In this paper, both processes will be implemented.
 
As described in the section \ref{relatedwork}, there is a considerable research body for various components of the mobile phones, but there are no reported research findings for the fingerprints of the magnetometers in a mobile phone. This paper aims to address this gap. 
\\
\\
The structure of this paper is the following: section \ref{relatedwork} provides an overview of the literature of electronic device fingerprinting in mobile phones. Section \ref{method} describes the methodology. In particular, the section describes the approach used to stimulate the magnetometers of the mobile phone in order to generate repeatable digital outputs. Then, the section describes the extraction of the statistical features from the digital output and the application of the \ac{SVM} machine learning algorithm to perform the classification and verification process. Then, section \ref{results} presents the results of the application of the classification algorithm on a set of nine phones of different models and brands including phones of the same model to test the intra-model identification. Finally, section \ref{conclusions} concludes the paper.  

\section{Related Work}
\label{relatedwork}
The fingerprint concept has been applied to different components of the mobile phones. There is an extensive literature on the possibility to fingerprint the digital camera of a mobile phone thanks to the imperfections present in the various components, which are part of the camera: lenses, \ac{CFA}, the sensor or even the software to process the images before storing them (e.g., compression algorithms). The \ac{SPN} presented in the pioneer work by \cite{lukasspn} is the most adopted and referenced approach because of its high accuracy even for intra-model identification and its persistence in time. The \ac{SPN} is based on the non-uniformity of each sensor pixel sensitivity to light. In comparison to other types of noise or imperfections having random distribution, the sensor pattern noise is a deterministic component, which stays the same for different images (in fact, it is strengthened with an increasing number of images). Even a limited set of 30-50 images can provide a \ac{SPN} with good accuracy \cite{Fridrich}.

Fingerprinting of the radio frequency components of a mobile phone has also been performed by various authors for different wireless standards. Researchers have applied \ac{RF} fingerprinting to 802.11a in \cite{suski2008using}, to \ac{GSM} in \cite{reising2010improved} and to ZigBee devices in \cite{Bihl}. In both cases, the fingerprinting technique is based on the selection of statistical features of the collected and processed radio frequency signals emitted by the mobile phone. The statistical features are variance, standard deviation, skewness and kurtosis (e.g., \cite{Bihl}), which are also used in this paper. Other statistical features based on the Hilbert-Huang Transform are used in \cite{Hilbert} for the \ac{GSM} standard. \ac{RF} fingerprinting usually provides very high accuracy (more than 90\%) both for intra-model and inter-model classification).

The microphone of a mobile phone is another component that can be used for identification and classification. In \cite{Hanilci}, the authors use \ac{MFCC} as fingerprinting feature because it is commonly employed as a feature to characterize human speakers in audio recordings. In \cite{Hanilci}, the authors apply \ac{MFCC} to the identification of the brand and model of a mobile phone on a experimental set of 14 different mobile phones. Apart from two mobile phones, which are of the same model and brand, all the other phones have different brands and models, so the analysis is mostly for inter-model rather than intra-model verification and identification. The authors used the \ac{SVM} classifier for identification and verification.

Another set of features for microphone fingerprinting was used in \cite{Kotropoulos}, where large-size raw feature vectors are obtained by averaging the log-spectrogram of a speech recording along the time axis for each mobile phone. The authors focused on inter-model identification and verification as they used mobile phones of various models from different brands.  The features were then fed to three distinct classifiers: the \ac{SRC}, the \ac{SVM} and \ac{NN}. \ac{SVM} provided the best performance in many configurations.

Finally, accelerometers and gyroscopes can also be used for fingerprinting as demonstrated in \cite{AccelPrint2014}, \cite{MobID_Stanford2014} and \cite{baldini2016experimental}. The mobile phones are submitted to repeatable motion patterns, which stimulate the accelerometers and gyroscopes to produce specific responses. The small variations in the responses are used to create the fingerprints and distinguish between the mobile phones. Both inter-model and intra-model identification and verification were performed with very good accuracy. \ac{SVM} and other classifiers were used in the cited references. 

While there is an extensive literature on the identification and verification of mobile phones for different types of components, there is no reported study on the identification through magnetometers. The goal of this paper is to address this gap. We note from this survey that the use of statistical features in combination with \ac{SVM} has been widely used in literature, which supports our choice to use a similar approach. 
 
\section{Methodology}\label{method}

\subsection{Workflow}
\label{workflow}
The goal of this section is to describe the overall methodology to perform the classification of the magnetometers built in the mobile phones, the test setup to collect the digital output (i.e., observables) of the magnetometers, the set of statistical features used to generate the fingerprints and the \ac{SVM} machine learning algorithm used to perform the classification.  

The overall workflow is presented in figure \ref{workflow} and the single steps are described in the following subsections.

\begin{figure*} [!ht]
	\centering
		\includegraphics[scale=0.4]{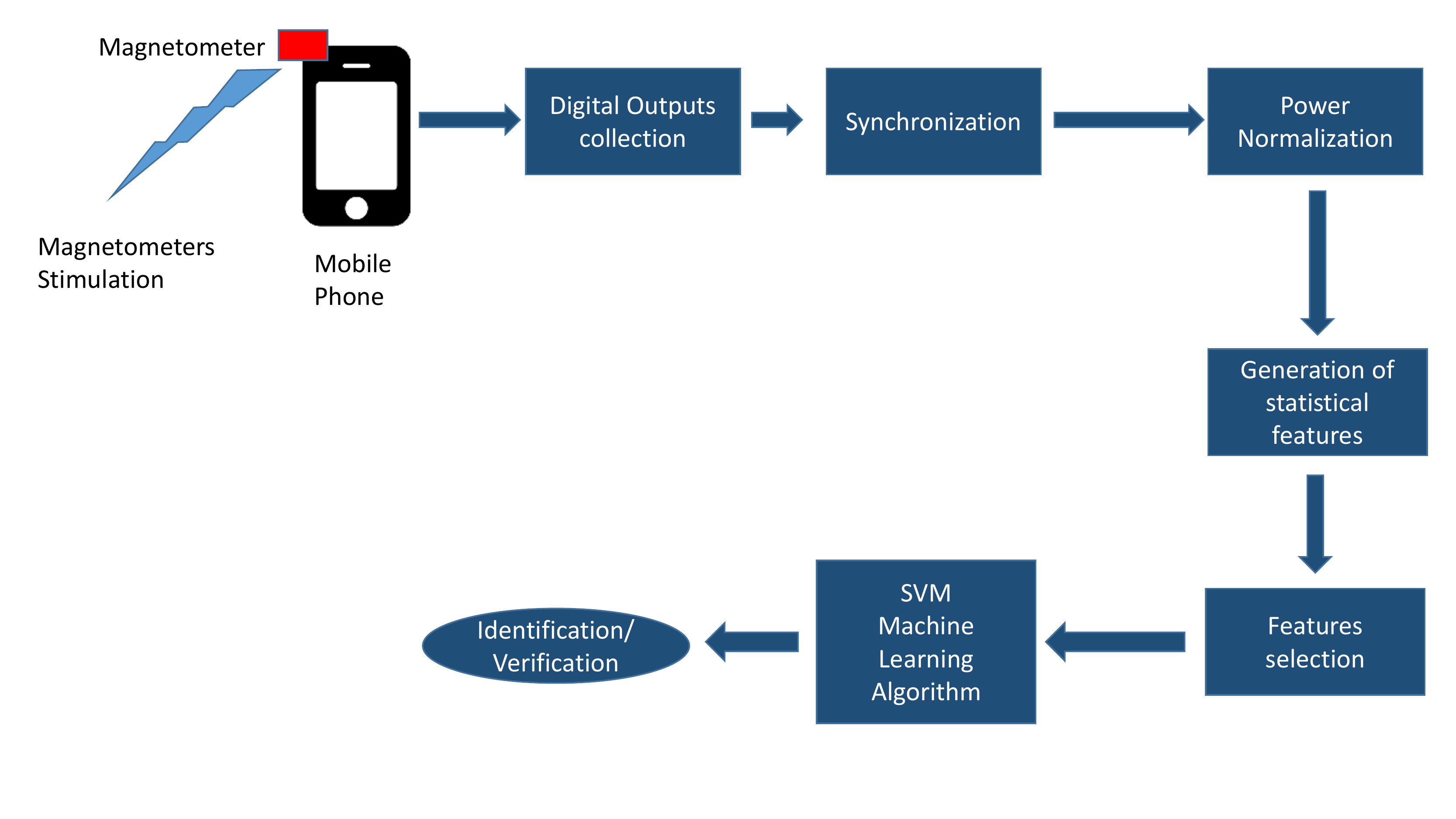}
	\caption{Workflow for the classification of mobile phones based on magnetometers fingerprints}
	\label{fig:methodology}
\end{figure*}

\subsection{Stimulation of the magnetometers}
\begin{figure} [!ht]
	\centering
		\includegraphics[width=\linewidth]{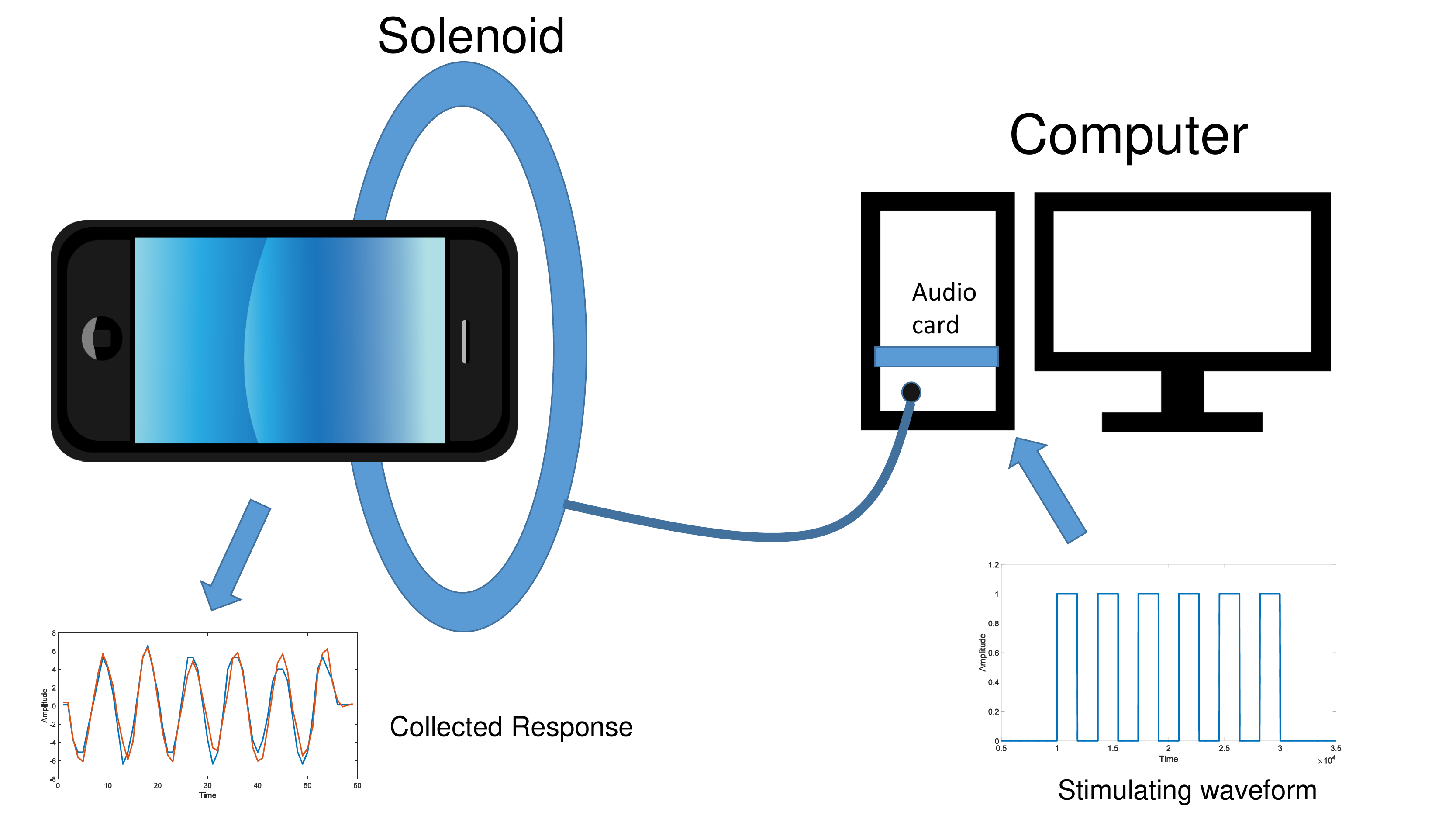}
	\caption{Test set-up to stimulate the magnetometers of a mobile phone}
	\label{fig:solenoid}
\end{figure}

\begin{figure} [!ht]

\begin{subfigure}{.5\textwidth}
  \centering
  \includegraphics[width=.8\linewidth]{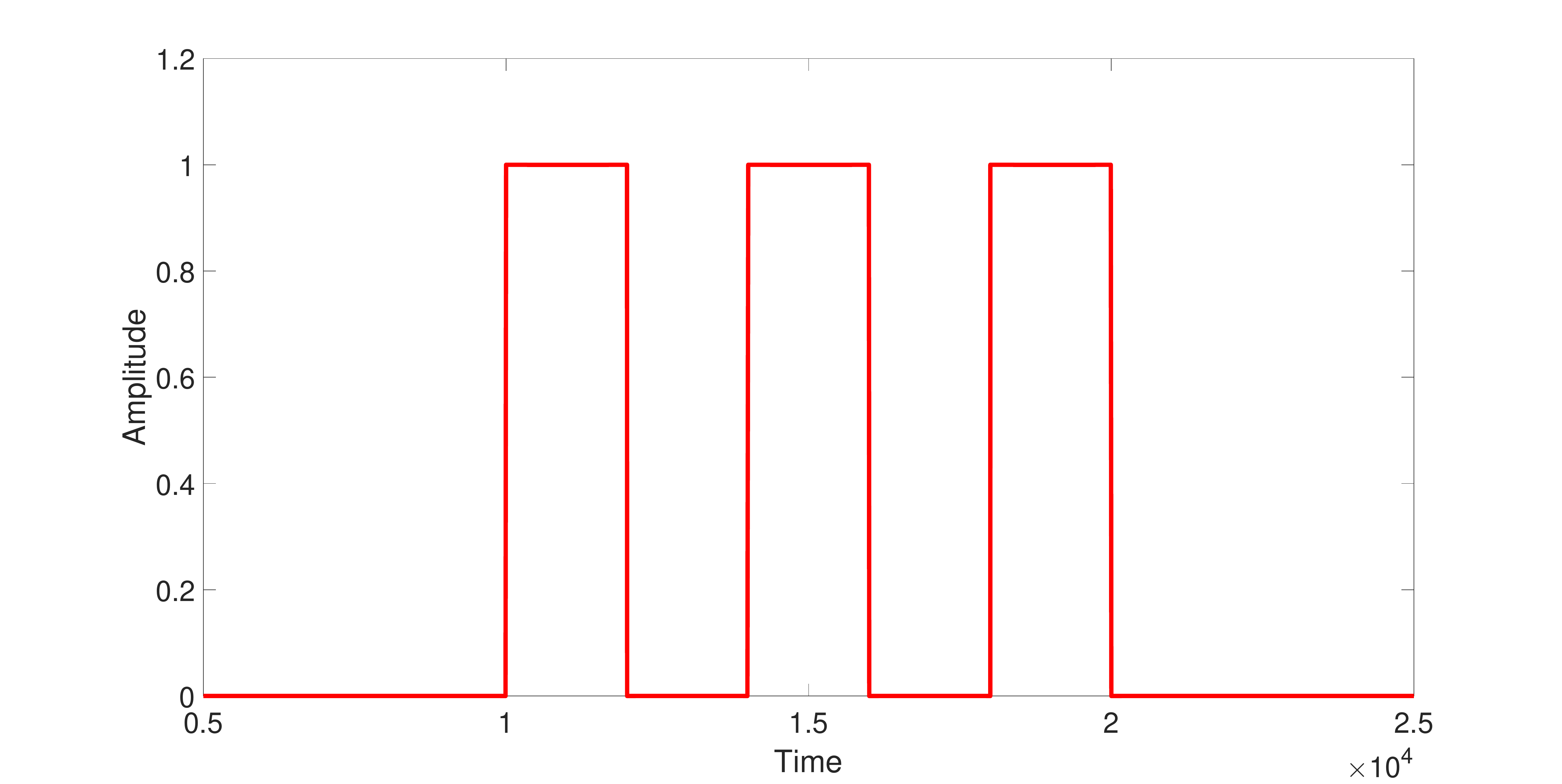}
  \caption{Waveform A (short)}
  \label{fig:stima}
\end{subfigure}

\begin{subfigure}{.5\textwidth}
  \centering
  \includegraphics[width=.8\linewidth]{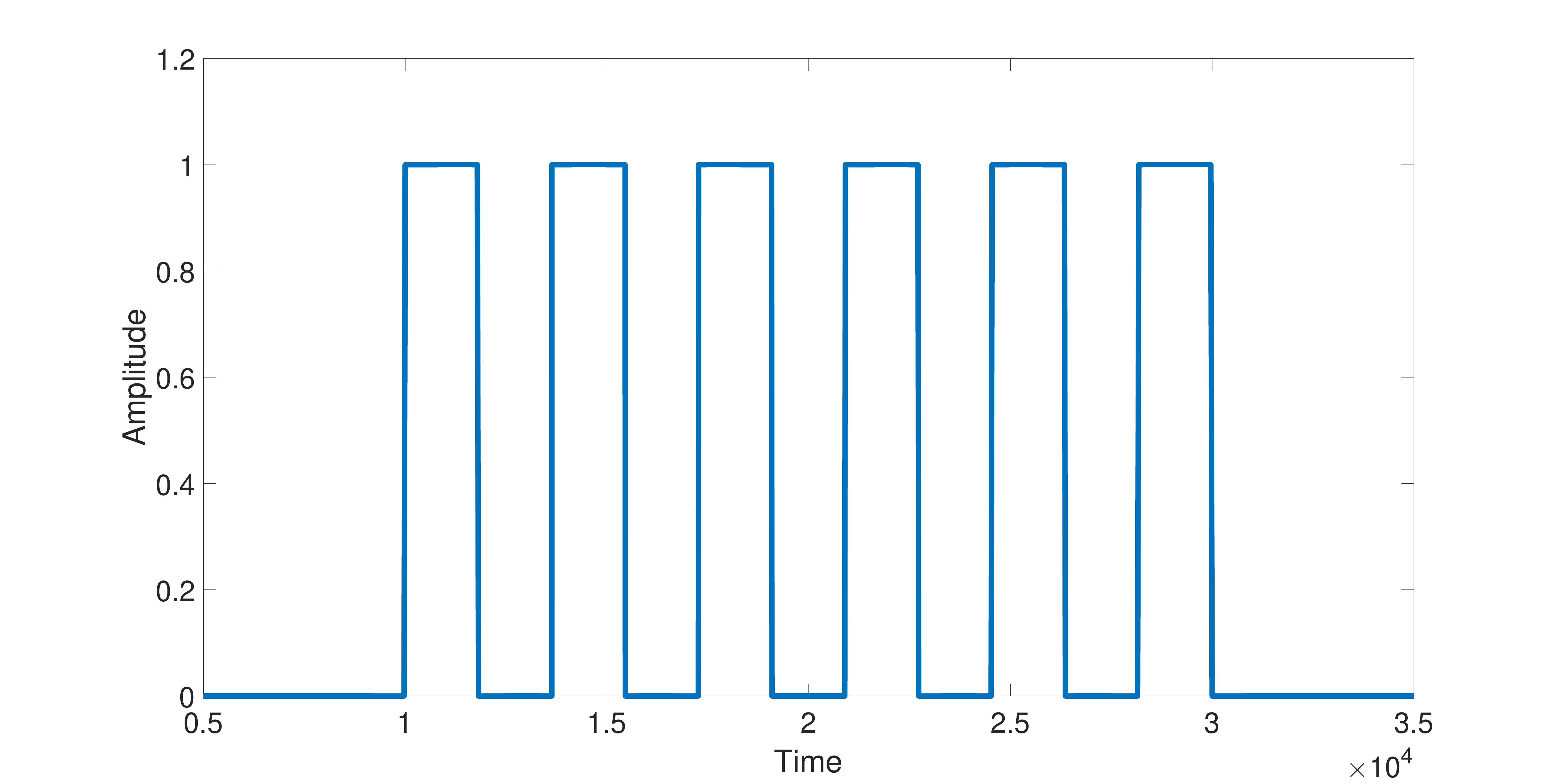}
  \caption{Waveform B (medium)}
  \label{fig:stimb}
\end{subfigure}

\begin{subfigure}{.5\textwidth}
  \centering
  \includegraphics[width=.8\linewidth]{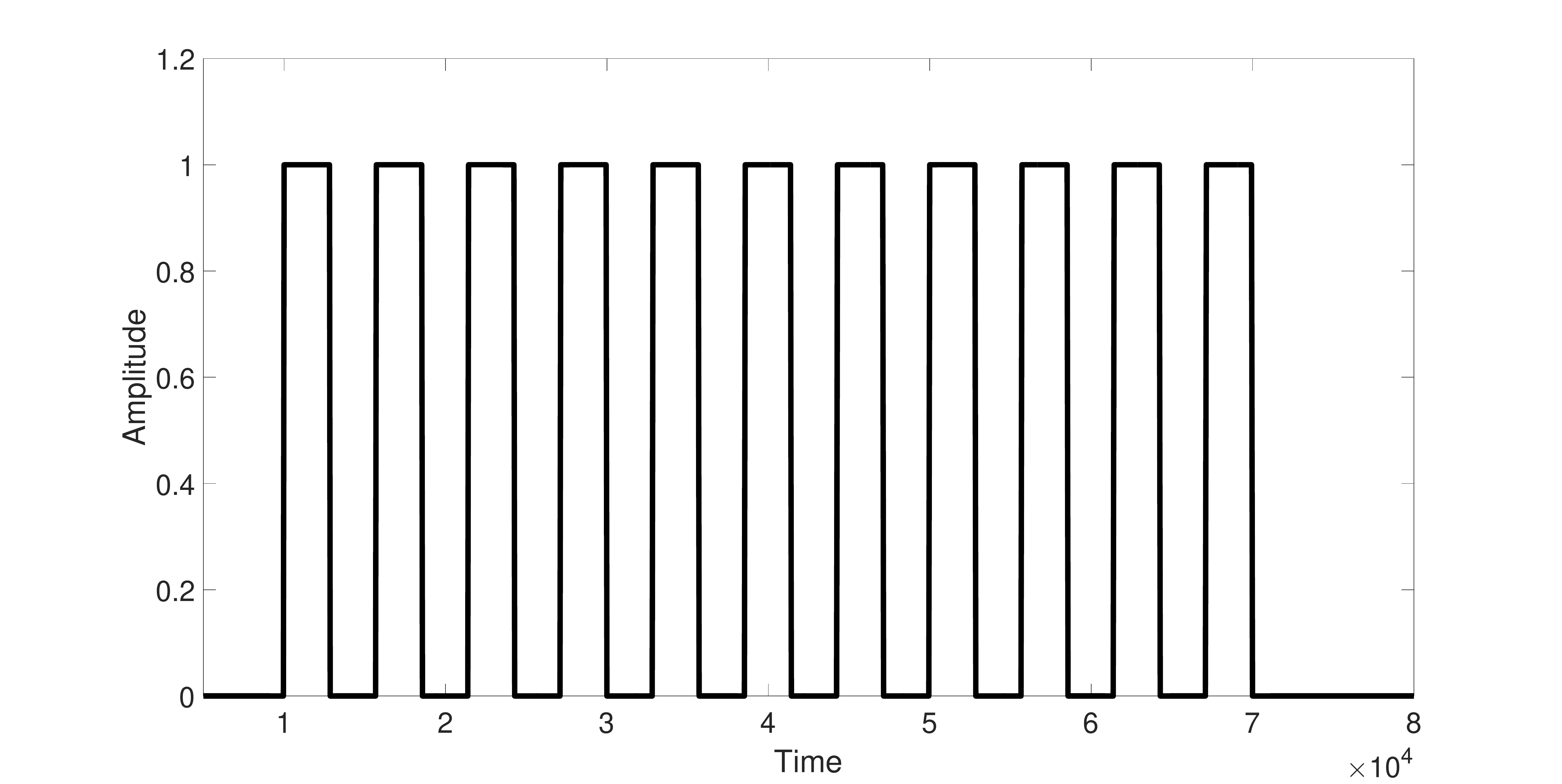}
  \caption{Waveform C (long)}
  \label{fig:stimc}
\end{subfigure}

\caption{Waveform patterns used to stimulate the mobile phones}
\label{fig:waveforms}
\end{figure}

The magnetometers in the phones are stimulated using a cost-effective solenoid, which is connected to the audio board of the computer. Since the goal of the experiment is to show the feasibility of mobile phone identification using low cost equipment, a consumer mass market audio board is used to stimulate the magnetometers because.
A picture of the schema used to stimulate the magnetometers of the mobile phone is shown in figure \ref{fig:solenoid}. The audio board runs a set of three synthetic waveforms based on square waves. The waveforms are shown in figure \ref{fig:waveforms}. The central waveform is the reference waveform used to conduct most of the measurements; the other two are used to show the impact on performance accuracy if the density of the square waves increases or decreases. In the rest of the paper, we will identify with waveform A the first waveform (figure \ref{fig:stima}), with waveform B the second waveform (figure \ref{fig:stimb}) and waveform C the third waveform (figure \ref{fig:stimc}). The waveforms were defined on the basis of the following considerations: a) a sharp impulse is needed to stimulate the magnetometers to generate adequate fingerprints, b) the distance between the square waves is defined on the average hysteresis values of the common mass market magnetometers and then empirically tested, c) a sequence of square waves generates response shapes by the mobile phones, which are appropriate for the use of the statistical features (e.g., skewness and kurtosis) defined in a subsequent section of this paper. The generated audio form is then played on the audio card connected to the solenoid. Note that the sharp rise of the square wave is what generates the impulse to the magnetometer.

Each phone is stimulated with a repetition of the 260 square waves described in \ref{fig:waveforms}. The value was chosen as a trade off between the need to have a statistical significant number of samples (a sample is the response of a mobile phone for a single waveform) and the time requested to stimulate the magnetometers (it is roughly one hour for the waveform B). 

\subsection{Digital outputs collection} 
\begin{figure} [!ht]

\begin{subfigure}{.5\textwidth}
  \centering
  \includegraphics[width=.8\linewidth]{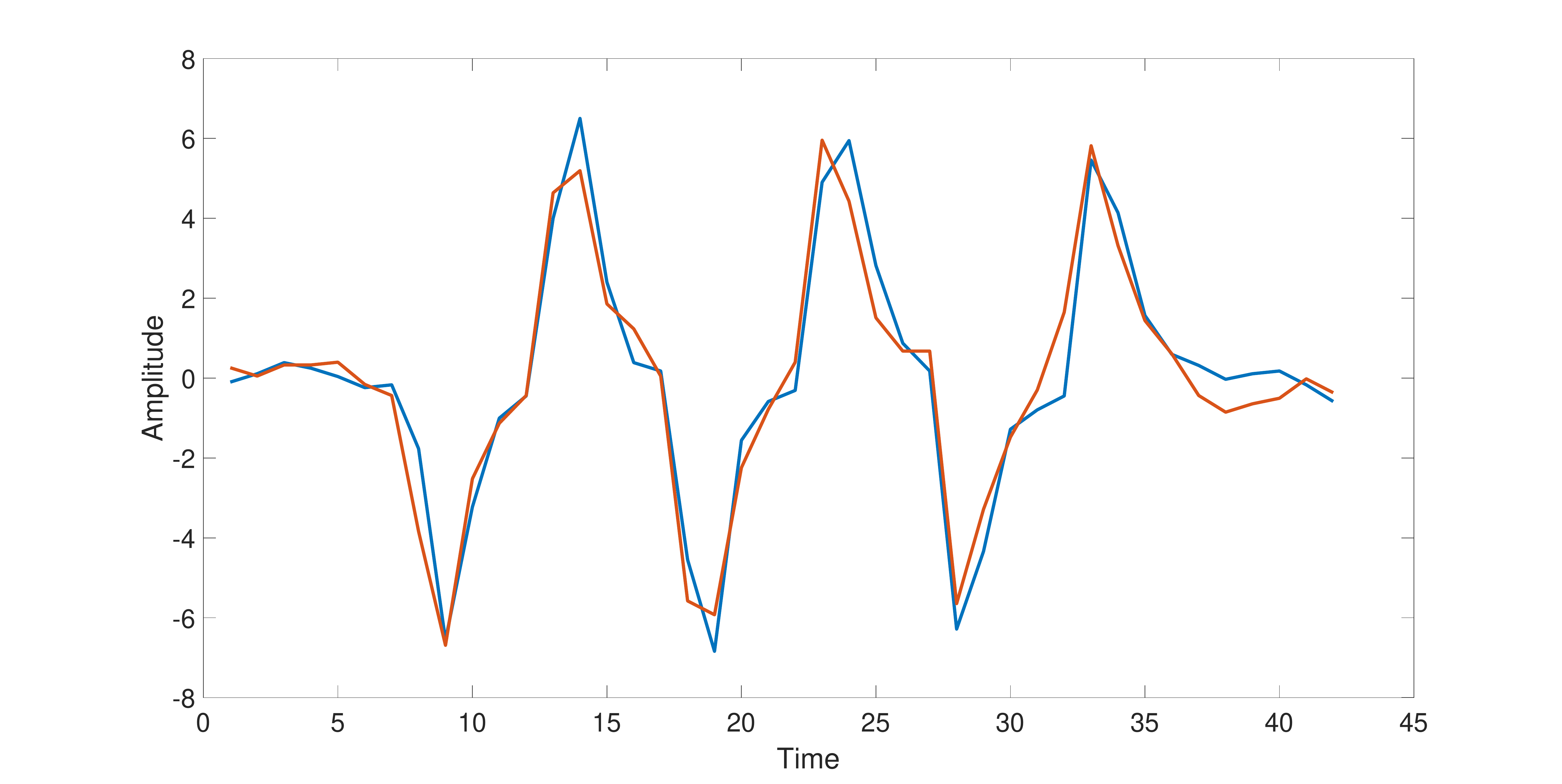}
  \caption{Response to Waveform A}
  \label{fig:respa}
\end{subfigure}

\begin{subfigure}{.5\textwidth}
  \centering
  \includegraphics[width=.8\linewidth]{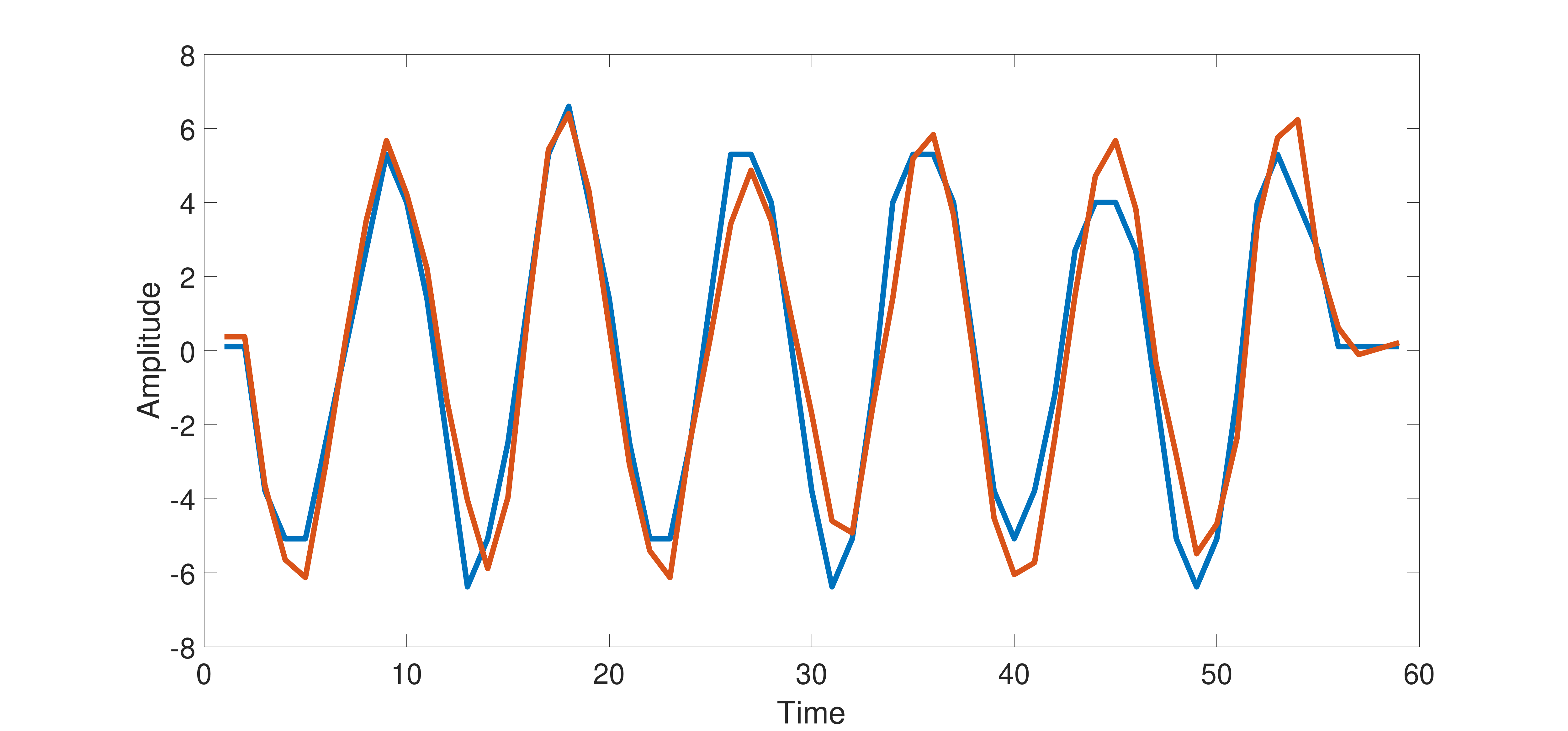}
  \caption{Response to Waveform B}
  \label{fig:respb}
\end{subfigure}

\begin{subfigure}{.5\textwidth}
  \centering
  \includegraphics[width=.8\linewidth]{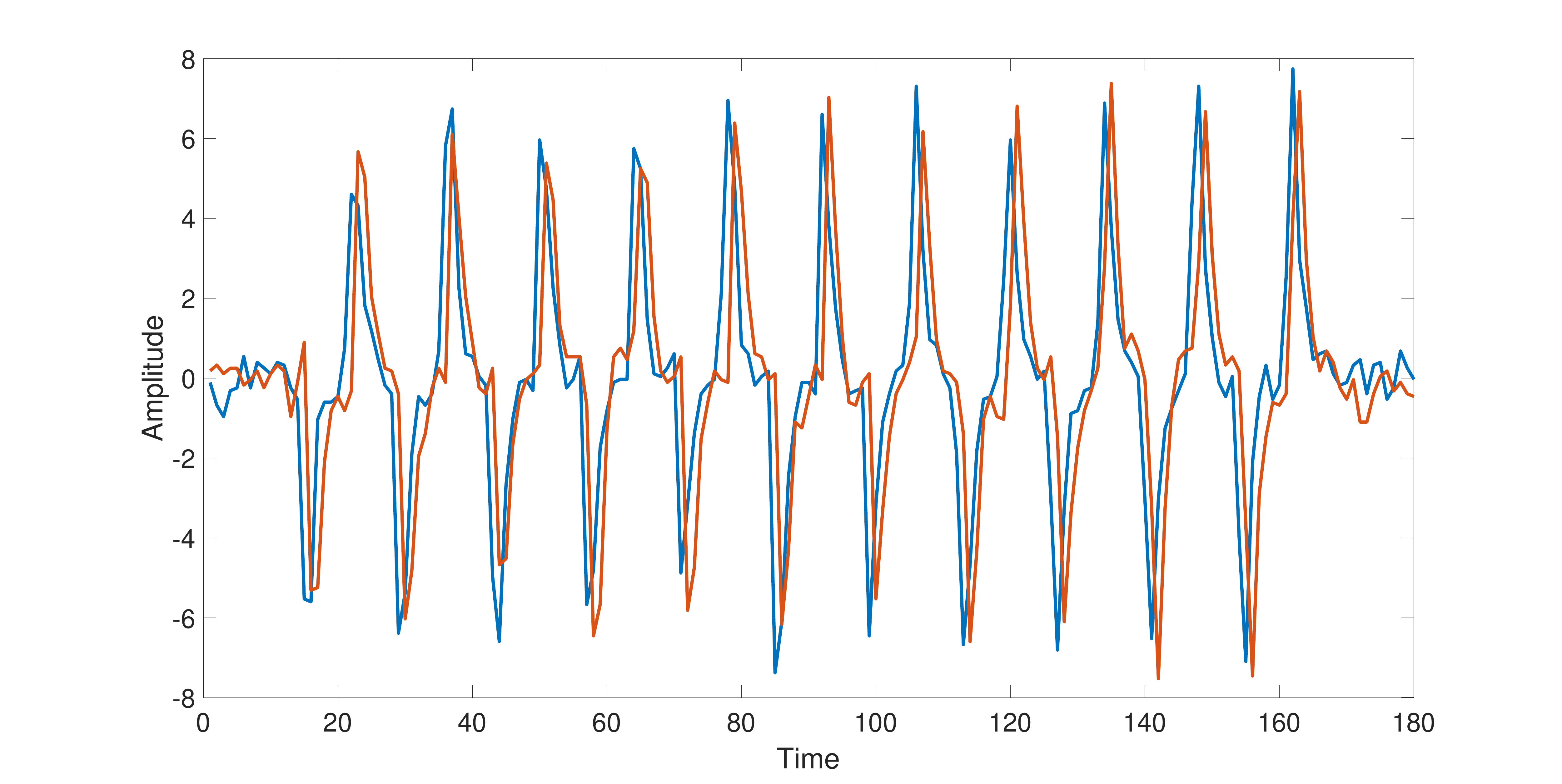}
  \caption{Response to Waveform C}
  \label{fig:respc}
\end{subfigure}

\caption{Responses collected by the mobile phones for different waveforms}
\label{fig:waveresponse}
\end{figure}

The digital output from the magnetometers is collected using a free available application called AndroSensor that we installed on the phones. In this experiment, we use only phones supporting the android operating system but a similar study can be performed on iOS based phones. The responses from two different mobile phones to the three waveform after synchronization and normalization is shown in figure \ref{fig:waveresponse}. We note that the response is quite manifold for different models of phones, which provides the unique fingerprint of the magnetometer (and the mobile phone) as described in section \ref{results}.
The collection of the data from the mobile phones was set to a frequency of 20 Hz: samples were taken by the magnetometers every 50 ms. This value was chosen because it was the lower common limit for all the mobile phones' sensors. 

The list of the 9 mobile phones used in the experiments is shown in table \ref{mobilePhones}. Phones are from 4 different brands and 4 models. The models for which more than one phone was used are HTC One (three devices), Samsung S5 (three phones) and Sony Experia (two devices). Although the three Samsung devices are of the same model, one of them runs a different version of the Android operating system. The selection of the Samsung mobile phones with different software version was done on purpose to evaluate the impact of different software. In most cases, mobile phones were produced in different years even if they are from the same model.

\begin{table}
	\centering
		\begin{tabular} {l | r}
		\textbf{Mobile phone model} & \textbf{Number of devices}\\ \hline
		HTC One & 3 \\ 
		Huawei Ascend Mate & 1 \\ 
		Samsung Galaxy S5 Android version 4.4 & 1 \\ 
		Samsung Galaxy S5 Andriod version 6.0 & 2 \\ 
		Sony Experia & 2 \\ \hline
		\end{tabular}
	\caption{List of Mobile Phones used in the experiment}
	\label{mobilePhones}
\end{table}

\subsection{Synchronization and Normalization}
The responses must be synchronized and normalized, otherwise the statistical features will generate false values, which are not dependent on the mobile phone itself but on other factors like the distance of the mobile phone from the solenoid. Synchronization and normalization is a common approach in fingerprinting (see \cite{suski2008using}, \cite{AccelPrint2014} and \cite{reising2010improved}).
The synchronization is performed using the Variance Trajectory technique. This technique is based on the calculation of the variance on a sliding window of samples, which moves along the response. The variance will increase substantially when the sliding windows meet a sharp rise or fall of the response. The rise of the variance identifies the beginning and the end of the response. This process is applied to all the 260 responses obtained in collection phase. The application of the variance trajectory was inspired by its use in \ac{RF} fingerprinting to detect the start and end of the wireless communication bursts (see \cite{suski2008using}).

The normalization is performed by applying the \ac{RMS} to each single response for each single mobile phone. 

\subsection{Statistical features}
\label{statfeatures} 
In this section, we define the statistical features used in this paper to generate the fingerprints. As described in \cite{Fulchertimeseries}, the classification of time series can be based on the representation of the time series using a set of derived properties, or features (i.e., feature-based classification). The advantage of the \textit{feature-based} approach is that it transforms a temporal problem in a static one and the verification and identification can be therefore more computationally efficient once the statistical features have been generated. Another advantage is the presence of many statistical features, which could be used for classification. This provides a larger set of tools for identification even if there is the risk that some statistical features are similar or correlated and increasing the number of features does not provide a significant increase in accuracy. Thus a features extraction or features selection process is needed. The disadvantage of the feature based approach is that it usually requires a statistical significant number of observables from the mobile phone to perform an accurate identification and verification. This is the reason why the magnetometers must be stimulated a significant number of times (260 in our case) to support the subsequent application of the machine learning algorithm.

Since this is the first attempt in literature to fingerprint magnetometers, there is no previous research work that can recommend a set of statistical features. Then, we use the references presented in the related work section to collect features, which are used by other authors. From the fingerprint work on \ac{RF} frequency components \cite{Bihl}, \cite{Reisingreduced}, we selected variance, standard deviation, skewness and kurtosis; from the work on accelerometers \cite{baldini2016experimental}, we selected entropy based features. 

The definition of the features is provided in equations (1) to (6): 
 
\begin{equation}
\label{eqn_entfeat}
\mbox{H}_{Shannon Entropy}\left\{ S_{TD} \right\} = -\sum_{i=1}^N {(S_{TD}^2*Ln (S_{TD}^2))}
\end{equation}

\begin{equation}
\label{eqn_logfeat}
\mbox{H}_{Log Energy}\left\{ S_{TD} \right\} = \sum_{i=1}^N {Ln (S_{TD}^2)}
\end{equation}

\begin{equation}
\label{eqn_stdfeat}
\mbox{Standard Deviation}\left\{ S_{TD} \right\} = \sqrt{\frac{1}{N-1} \sum_{i=1}^N {(S_{TD}-\mu)^2}}
\end{equation}

\begin{equation}
\label{eqn_varfeat}
\mbox{Variance}\left\{ S_{TD} \right\} = \frac{1}{N-1} \sum_{i=1}^N {(S_{TD}-\mu)^2}
\end{equation}

\begin{equation}
\label{eqn_skewfeat}
\mbox{Skewness}\left\{ S_{TD} \right\} = \frac{1}{\sigma^3} \sum_{i=1}^N {(S_{TD}^3-{\mu}^3)}
\end{equation}

\begin{equation}
\label{eqn_kurfeat}
\mbox{Kurtosis}\left\{ S_{TD} \right\} = \frac{1}{\sigma^4} \sum_{i=1}^N {(S_{TD}^4-{\mu}^4)},
\end{equation}

where the term $S_{TD}$ represent the instantaneous amplitude of the signal.

The response was also converted in the frequency domain with a \ac{FFT}. Then, the statistical features identified in equations (1) to (6) were applied respectively to the Phase and the Amplitude of the frequency domain representation of the response. 
This transformation provides a set of additional 12 features (6 for the phase of the FFT and 6 for the amplitude of the FFT) for a total of 18 features. The \ac{FFT} transformation was implemented because it is a common practice applied in the surveyed papers \cite{Hilbert} but also because the empirical evidence from the \ac{FFT} has shown subtle differences, which could be exploited for classification. 

In the rest of this paper and especially in the results section \ref{results}, we identify the features with the numbering of the equations: 1-6 in the time domain, 7-12 for the phase in the frequency domain and 13-18 for the amplitude in the frequency domain. 

\begin{figure} [!ht]
	\centering
		\includegraphics[width=\linewidth]{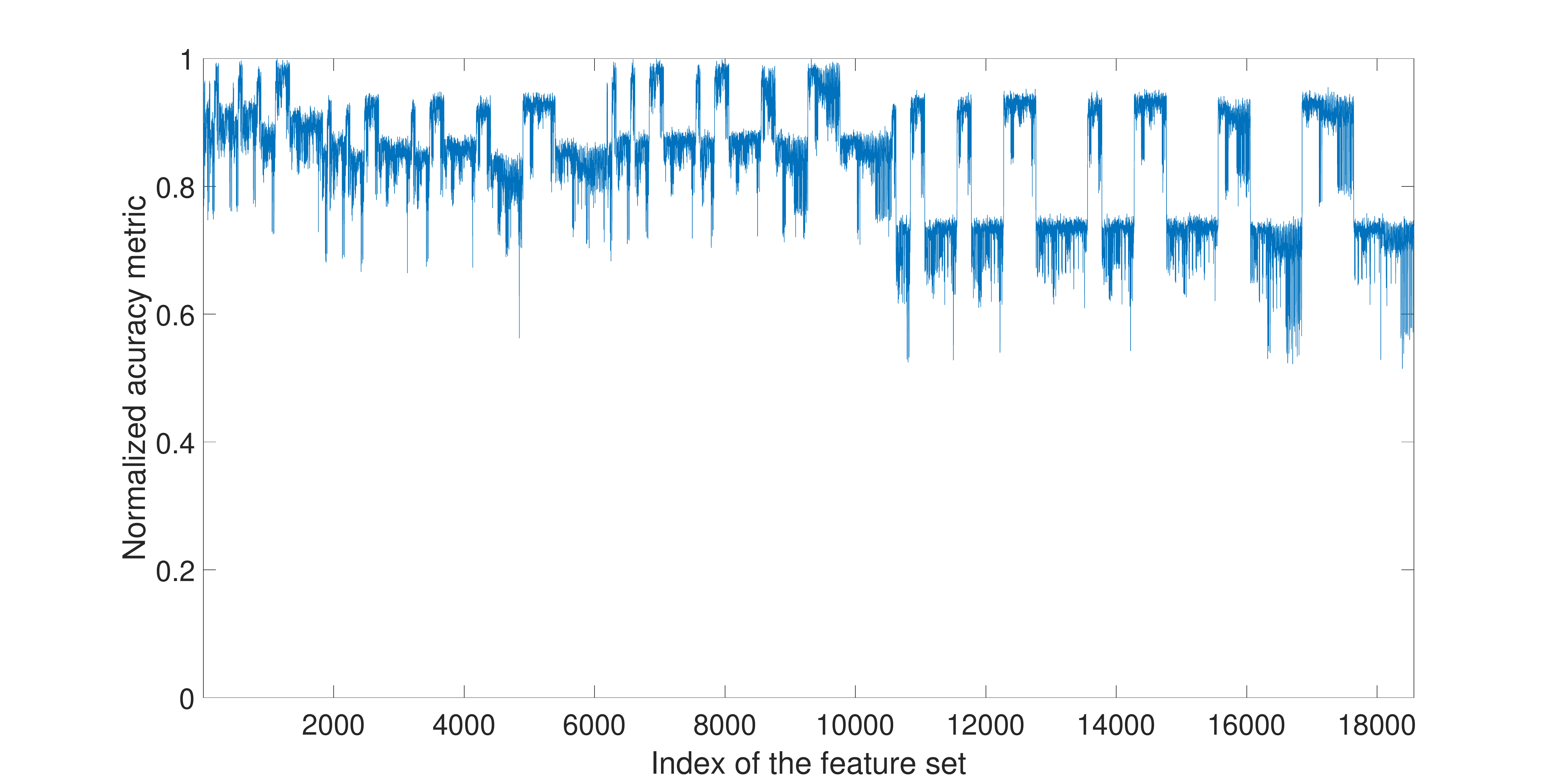}
	\caption{Accuracy metric against the set of features for one fold}
	\label{fig:plotfeatures}
\end{figure}

Combinations of statistical features extracted from the output of magnetometers represent their fingerprints. The goal is to select the best combination of features that can provide the optimal identification and verification accuracy. The process to achieve this goal is called features selection. Various approaches have been proposed in literature for features selection (see \cite{Fulchertimeseries}). In this paper, we adopt a combination of a brute force approach with the \ac{SFS} algorithm. This algorithm starts with a single feature or a small set of features and incrementally adds a feature at the time and measures the resulting value of metric. If the metric improves, the feature is added, otherwise another feature is checked. The process continues until the maximum (i.e., optimal) value of the metric is reached. In this paper, a metric based on the accuracy was used for the \ac{SFS} algorithm. The accuracy is described in section \ref{smvopt}.
To avoid local maximum values, a brute force approach is also performed to select one or few sets of combinations of 6 features among all the possible permutations of the features (groups of 6 on a total set of 18 features, which results in 18564 features). In the brute force approach, all the possible combinations of 6 features were calculated. The process was repeated for all the folds used in the machine learning classification (see \ref{smvopt}). An example for one fold is provided in the normalized graph of figure \ref{fig:plotfeatures} where the presence of peaks (maximum values) is evident. From these optimum values, the  \ac{SFS} algorithm calculated which features should be added. The final result of the example of a single fold is that the best combination of features is [1 2 3 6 7 10 15 17] (where two additional features were added to one of the best sets identified from the brute force approach).

\subsection{Machine Learning algorithms and parameters optimization}
\label{smvopt}

The \ac{SVM} is a very well know technique in supervised machine learning and it has been used in this paper since it \ac{SVM} has demonstrated its effectiveness for \ac{RF} fingerprinting in the related research literature (see \cite{hasse2013forensic}, \cite{Hanilci} and \cite{baldini2016experimental}.

On the basis of a set of training samples (in the case study presented in this paper, the training samples are the features of the magnetometers responses), the \ac{SVM} algorithm assigns each sample to one of two categories in the training phase. This makes \ac{SVM} a non-probabilistic binary linear classifier. The resulting \ac{SVM} model is a representation of the samples as points in space, mapped so that the samples of the separate categories are divided by a clear gap that is as wide as possible. In the testing phase (e.g., the verification of the identify of a mobile phone) new samples are then mapped into that same space and predicted to belong to a category based on which side of the gap they fall on.

Here, we provide a brief formal description of the \ac{SVM} algorithm to highlight the key points, which will be used in our study. Additional details on the \ac{SVM} algorithm can be found in \cite{cristianini2000introduction}.

We define the $n$ labeled examples $(x_{1},y_{1}),\ldots, (x_{n},y_{n})$ with labels $y_{i}\in \{1,-1\}$. We want to find the hyperplane H (in a proper d-dimensional space K) defined by $<w,x>+b = 0$ (i.e. with parameters $(w,b)$ ), which satisfies the following conditions:

\begin{enumerate}
  \item The scale of $(w,b)$ is fixed so that the H plane is in canonical
        position,
\[ \min_{i\le n} | <w,x_{i}> + b | = 1 \]

  \item The H plane with parameters $(w,b)$ separates the $+1$'s from the
        the $-1$'s. as described from the following equation, 
\[ y_{i} ( <w,x_{i}> + b ) \ge 0 \mbox{\ for all $i\le n$} \]

  \item The H plane has a maximum margin $\lambda = 1/|w|$. i.e., minimum
        $|w|^{2}$.
\end{enumerate}

We can redefine $<w,x>+b = 0$ to the following equation:

    \begin{equation}
      w \bullet \phi(x) + b =0
    \end{equation}  

where  $\phi (x)$ represents a proper mapping of x into the space K; $w = [w_1;w_2; :::;w_d]$ represents a d-dimensional real vector
normal to H and b is a real parameter such that $|b|/||w||$ is the perpendicular distance of the origin from H. $\bullet$ is the scalar product between two vectors.

The problem of finding the hyperplane H is an optimization problem, which can be defined on the basis of the following equation:

    \begin{equation}
      min_{(w,b,\upsilon)} \frac{1}{2} {W^T} {W} + C \sum \upsilon_{i}
    \end{equation}  

Where $\upsilon_{i}$ are the slack variables and \textit{i} is in the range of 1 to $N_{Train}$, which is the number of training vectors. The slack variables are subject to $\upsilon_{i}>0$ and they account for the presence of classification errors. The parameter C (which we will call Box Constraint in the rest of this paper) allows the SVM user to control the weight of the classification errors in the previous equation and it is one of the two parameters to be tuned in the training process.

The second parameter to be tuned is related to the Kernel function, which is used to define the shape and format of the hyperplane. Various kernel functions are available in literature including linear, polynomial and \ac{RBF}.

In this paper, we use \ac{SVM} with \ac{RBF} as a kernel function because it has demonstrated its effectiveness for fingerprinting classification in \cite{hasse2013forensic} and other references. In addition, this kernel has a number of good features, since it can properly handle the cases in which the relation between class labels and features is nonlinear in classification problems (which is indeed our case).

The definition of the \ac{RBF} is the following:

    \begin{equation}
      K(\mathbf x_i, \mathbf x_j) = e^{-{\gamma}\|\mathbf x_i - \mathbf x_j\|^2},
    \end{equation}  
		
where the $\gamma$ scaling factor is the second parameter to be tuned together with C Box Constraint parameter. 

To summarize, the application of \ac{SVM} in this context requires the optimization of the statistical features and the parameters of the \ac{SVM} and the \ac{RBF}, which are the scaling factor $\gamma$ from equation (10) and the Box Constraint C parameter from equation (9).

To avoid the problem of over-fitting in the training phase, where a single set of training data can contain bias not present in other data sets, a 10-fold partition is used for training and classification. In the 10-fold method, each collection of statistical fingerprints (one for each mobile phone) is divided into ten blocks. Nine blocks are used for training and one block is \textit{held out} for classification. The training and classification process is repeated ten times until each of the ten blocks has been \textit{held out} and classified. In this way each block of statistical fingerprints is used once for classification and nine times for training. The final cross-validation performance statistics are calculated by averaging the results of all folds. In this way, the presence of bias in a specific training set or in portions of the training set are averaged or mitigated.

\ac{SVM} is a binary classifier, and to perform classification of more than two systems as in our case (i.e., 9 mobile phones), we need to use a multi-classifier based on \ac{SVM}. Two common  approaches  are  the  \ac{OAO} and  \ac{OAA} techniques \cite{cristianini2000introduction}.  \ac{OAA} involves  the  division  of  an  N (i.e., 9 in our case) class data-sets  into  N  two-class  cases, while \ac{OAO} involves the creation of a classification machine composed by N(N-1)/2 machines for each pair of systems. While \ac{OAO} is more computationally intensive than \ac{OAA} (N(N-1)/2 against N), OAA has some disadvantages especially with unbalanced training data-sets. Since we have a limited set of systems (i.e., nine) and the observables are also in limited number (i.e., 260), performance is not an issue and we decided to select \ac{OAO}.

Finally, we adopted the \ac{SMO} for \ac{SVM}. \ac{SMO} is an algorithm for solving the quadratic programming (QP) problem and it is commonly used in machine learning.

The Sequential forward selection algorithm must be based on a criterion against which the optimum value is identified. 
In machine learning, the following parameters are defined:
\begin{itemize}
	\item $T_p$ is the number of true positive matches where the machine learning algorithm has correctly identified a sample (e.g., a collected \ac{RF} signal in our context) as belonging to the correct class.
	\item $T_n$ is the number of true negative matches where the machine learning algorithm has correctly identified a sample as not belonging to the correct class.
	\item $F_p$ is the number of false positive matches where the machine learning algorithm has identified a sample as belonging to a class while it is not true.
	\item $F_n$ is the number of false negative matches where the machine learning algorithm has identified a sample as not belonging to the class while this is not true. 
\end{itemize}

The combination of the different parameters can define different metrics to evaluate the effectiveness of a machine learning algorithm. 
In this case, we use the verification accuracy as a criterion, which is calculated as:

\begin{equation}
Accuracy = \frac{T_p+T_n}{Total Population},
\label{eqn_acc}
\end{equation}

\noindent where $T_p$ is the number of true positives and $T_n$ is the number of true negatives resulting from the application of the \ac{SVM} machine learning algorithm to the problem of verifying that the collected fingerprints are representative of the same magnetometer evaluated in the training phase (i.e., for verification). The total population represents the total population of samples (which is the sum of $T_p$, $T_n$, $F_p$ and $F_n$). 

Beyond accuracy, in this paper, we will also use \ac{ROC} and the \ac{EER} metrics to evaluate the verification accuracy.

The \ac{ROC} is calculated as \ac{ROC}-like performance curve and it is generated by plotting the $F_p$ vs. $F_n$ as the verification threshold changes.

The \ac{EER} corresponds to the point on the \ac{ROC} curve where the $F_p$ vs. $F_n$ are equal. This metric is frequently used as a summary statistic to compare the performance of various classification systems. In general, a lower \ac{EER}s indicate better system classification performance.

Finally, the confusion matrix is also used to show the results of the identification process. In the confusion matrix, each column of the matrix represents the instances of a predicted class while each row represents the instances of the actual class (and vice-versa). As in our experiments we used 9 phones, the confusion matrix shown in the results section \ref{results} has a dimension of 9*9. In the confusion matrix, the correct guesses (i.e., true positive or negative) are located in the diagonal of the table, so it's easy to inspect the table for errors, as they will be represented by values outside the diagonal. 
The confusion matrix is also used in this paper to define the metric employed in the \ac{SFS} and for the optimization of the scaling factor and box constraint parameters. The metric is the sum of the diagonal values of the confusion matrix divided for all the values of the confusion matrix. It can be considered an extension of the accuracy metric. In general, the confusion matrix is used for classification purposes while the \ac{ROC}s are used for verification.

Using the accuracy metric derived from the confusion matrix, we calculated the best set of features: [1 2 3 6 7 10 15 17] as described before and then we identified the optimum values of the scaling factor and the box constraint. This was achieved by creating a two dimensional array based on a set of values of scaling factor ranging from $2^{-8}$ to $2^8$ and the box constraint ranging from $2^{-8}$ to $2^{28}$. Similar ranges of values are suggested by various references like \cite{cristianini2000introduction}.

\begin{figure} [!ht]
	\centering
		\includegraphics[width=\linewidth]{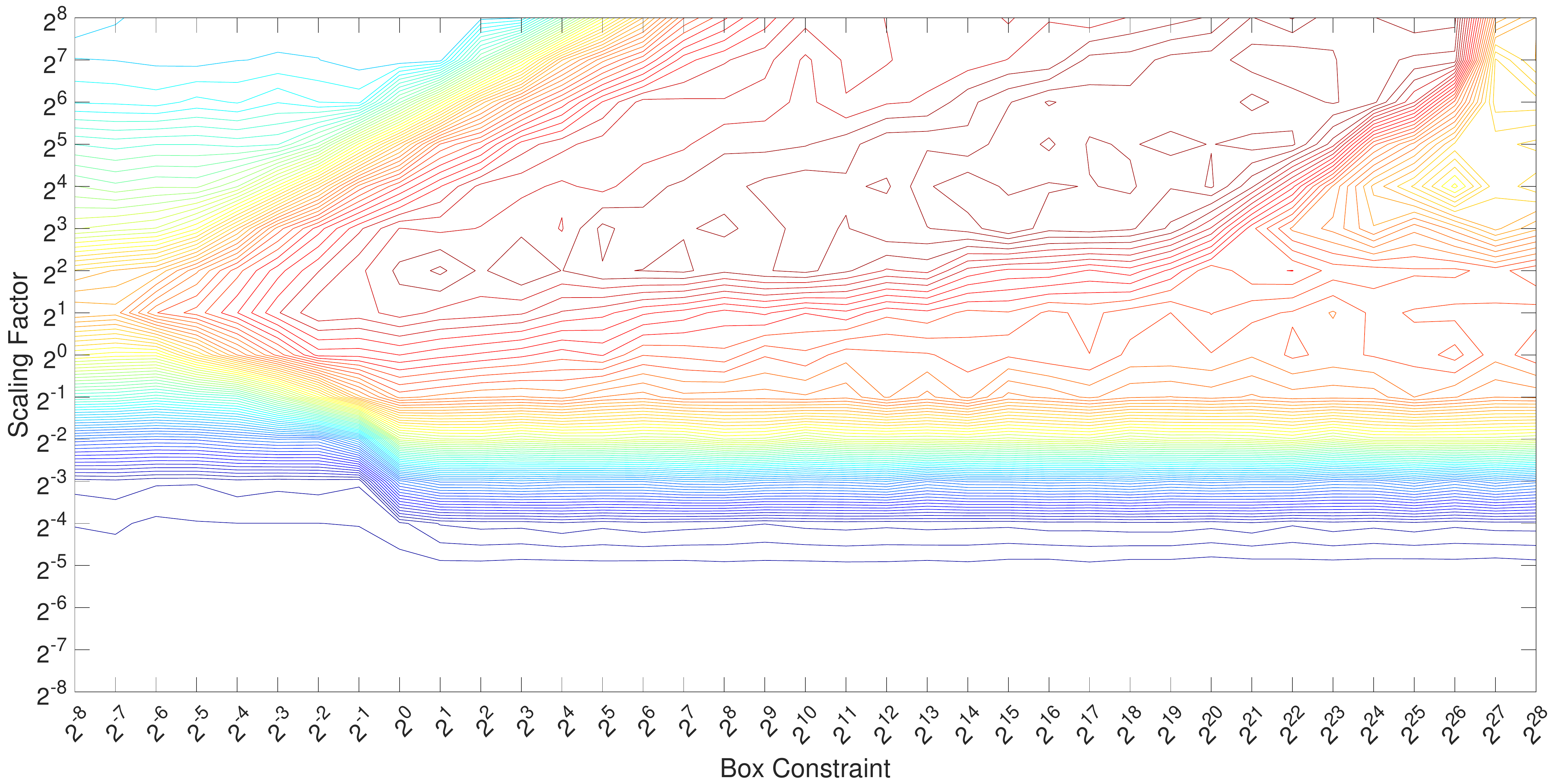}
	\caption{Optimization of the scaling Factor and box Constraint values for one fold}
	\label{fig:scalingbox}
\end{figure}

The result is shown in figure \ref{fig:scalingbox}, which features a peak value in correspondence of a Scaling Factor equal to $2^7 (128)$ and a Box Constraint of $2^{22}$ (4194304) for a specific fold as an example. 
The identified set of features and optimal values of the Scaling Factor and Box Constraint for each fold are used to produce all the graphs and results in the section \ref{results}.

\section{Results}
\label{results}

\subsection{Results based on one day of measurements}
In this section, we describe the experimental results for identification and accuracy.

Table \ref{tab:mobilePhones} is the confusion matrix of all the mobile phones stimulated with the Waveform B using the \ac{SVM} algorithm. From the table, we can immediately see that intra-model classification is very difficult to achieve with this approach, while inter-model classification is possible with a very good level of accuracy. More specifically, intra-brand classification is easy to achieve, because the algorithms do not confuse observables from different brands. The cross-values between HTC, Samsung and Sony are basically empty. Intra-model classification is difficult to achieve for the HTC One and SAMSUNG models, while we basically have random choice for the SONY Experia model. We also notice that the different versions of the operating systems (version 4 against version 6) for the SAMSUNG models do not have significant impact on the classification accuracy (i.e., no effects on data collection from magnetometers, which have the same characteristics).

\begin{table}
	\centering
	\resizebox{\columnwidth}{!}
	{
		\begin{tabular} {l|r|r|r|r|r|r|r|r|r}
		&
		\begin{turn}{90}HTC One 1\end{turn} &
		\begin{turn}{90}HTC One 2\end{turn} &
		\begin{turn}{90}HTC One 3\end{turn} &
		\begin{turn}{90}Huawei\end{turn} &
		\begin{turn}{90}SAMSUNG 1 (v4)\end{turn} & 
		\begin{turn}{90}SAMSUNG 2 (v6)\end{turn} & 
		\begin{turn}{90}SAMSUNG 3 (v6)\end{turn} & 
		\begin{turn}{90}SONY 1\end{turn} & 
		\begin{turn}{90}SONY 2\end{turn} \\ \hline
		HTC One 1 & 101 & 74 & 79 &	6 &	0 &	0	 & 0	& 0 & 0 \\ \hline
		HTC One 2 & 36 & 157 & 33 &	34 &	0 &	0	 & 0	& 0 & 0 \\ \hline
		HTC One 3 & 32 & 38 & 185 &	5 &	0 &	0	 & 0	& 0 & 0 \\ \hline
		Huawei & 0 & 29 & 0 & 231 &	0 &	0 &	0	 & 0	& 0 \\ \hline
		SAMSUNG 1 (v4) & 1 & 0 & 0 &	1 &	190 &	16	& 49	& 2 & 1 \\ \hline
		SAMSUNG 2 (v6) & 0 & 0 & 0 &	0 &	25 &	197	& 38	& 0 & 0 \\ \hline
		SAMSUNG 3 (v6) & 0 & 0 & 0 &	0 &	29 &	33	& 195	& 3 & 0 \\ \hline
		SONY 1 & 0 & 0 & 0 & 0 & 3 &	0	& 0	& 137 & 120 \\ \hline
		SONY 2 & 0 & 0 & 0 & 1 & 3 &	2	& 0	& 123 & 131 \\ \hline
		\end{tabular}
		}
	\caption{Confusion matrix for all the mobile phones using Support Vector Machine}
	\label{tab:mobilePhones}
\end{table}

We also evaluated the performance of the \ac{SVM} algorithm against the well known k-nearest neighbors algorithm (KNN). The resulting confusion matrix is shown in figure \ref{tab:mobilePhonesknn}, where it can be seen that the results are worst than what obtained with \ac{SVM}. This can also be evaluated by calculating the ratio of the sum of the diagonal elements between the two confusion matrix (the overall sum of the elements of the matrix is the same in both cases). The result is $1545 (SVM) \div 1361 (KNN) = 1.135$, which shows a clear advantage for \ac{SVM} (around 13\% better accuracy).

\begin{table}
	\centering
	\resizebox{\columnwidth}{!}
	{
		\begin{tabular} {l|r|r|r|r|r|r|r|r|r}
		&
		\begin{turn}{90}HTC One 1\end{turn} &
		\begin{turn}{90}HTC One 2\end{turn} &
		\begin{turn}{90}HTC One 3\end{turn} &
		\begin{turn}{90}Huawei\end{turn} &
		\begin{turn}{90}SAMSUNG 1 (v4)\end{turn} & 
		\begin{turn}{90}SAMSUNG 2 (v6)\end{turn} & 
		\begin{turn}{90}SAMSUNG 3 (v6)\end{turn} & 
		\begin{turn}{90}SONY 1\end{turn} & 
		\begin{turn}{90}SONY 2\end{turn} \\ \hline
		HTC One 1 & 92 & 71 & 76 &	19 &	0 &	0	 & 0	& 0 & 2 \\ \hline
		HTC One 2 & 47 & 124 & 29 &	60 &	0 &	0	 & 0	& 0 & 0 \\ \hline
		HTC One 3 & 52 & 51 & 143 &	14 &	0 &	0	 & 0	& 0 & 0 \\ \hline
		Huawei & 4 & 21 & 1 & 234 &	0 &	0 &	0	 & 0	& 0 \\ \hline
		SAMSUNG 1 (v4) & 1 & 0 & 1 &	 &	152 &	37	& 65	& 0 & 4 \\ \hline
		SAMSUNG 2 (v6) & 0 & 0 & 0 &	0 &	37 &	174	& 49	& 0 & 0 \\ \hline
		SAMSUNG 3 (v6) & 0 & 0 & 0 &	0 &	31 &	49	& 177	& 1 & 2 \\ \hline
		SONY 1 & 0 & 0 & 0 & 0 & 5 & 1	& 1	& 142 & 111 \\ \hline
		SONY 2 & 0 & 0 & 3 & 0 & 4 &	2	& 2	& 126 & 123 \\ \hline
		\end{tabular}
		}
	\caption{Confusion matrix for all the mobile phones using the k-nearest neighbors algorithm}
	\label{tab:mobilePhonesknn}
\end{table}

The results from the confusion matrix are confirmed by the ROC curves among the different phones of different brands, models or serial number of the different models.

\begin{figure} [!ht]
	\centering
		\includegraphics[width=\linewidth]{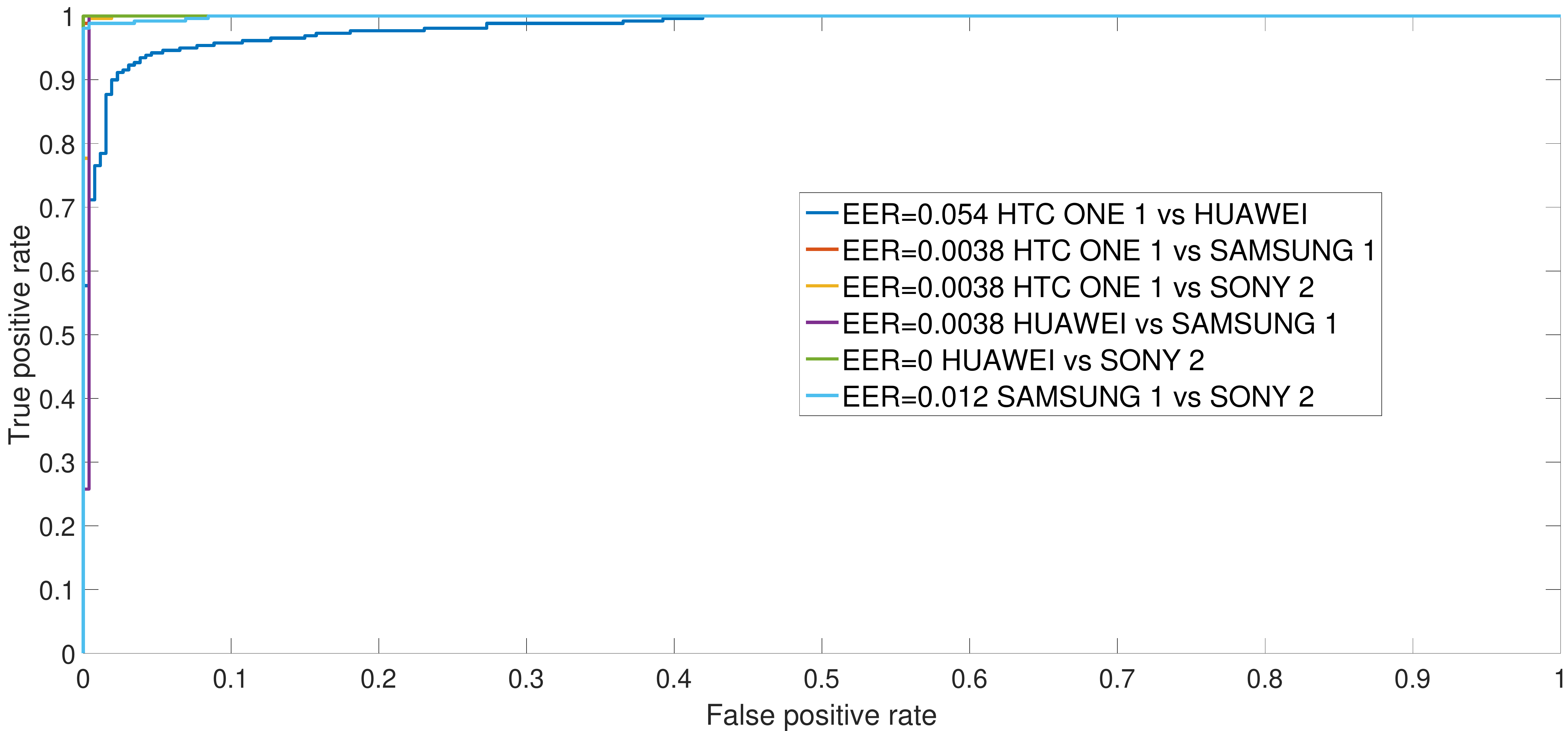}
	\caption{Inter-model and brand verification for day one}
	\label{fig:intermodelday1}
\end{figure}

Figure \ref{fig:intermodelday1} shows the \ac{ROC} curves and the related \ac{EER} values for verification among brands and models. ROCs are the results of a binary classification and they can be used for the verification process between a mobile phone A and B. In other words, if a mobile phone B falsely claims that it is actually mobile phone A, the algorithm should be able to verify the authenticity. ROC curves, which are more leaning against the center of the graph, indicate a lower verification accuracy. In fact, the \ac{EER} parameter is the intersection of the diagonal line with the ROC curve. The higher is an EER, the lower is the verification accuracy. 

We notice that it is very easy to distinguish between mobile phones of different brands (HTC against Huawei, Samsung or Sony) as expected from the confusion matrix. The EER values are very low, which demonstrate a very high verification accuracy.

\begin{figure} [!ht]
	\centering
		\includegraphics[width=\linewidth]{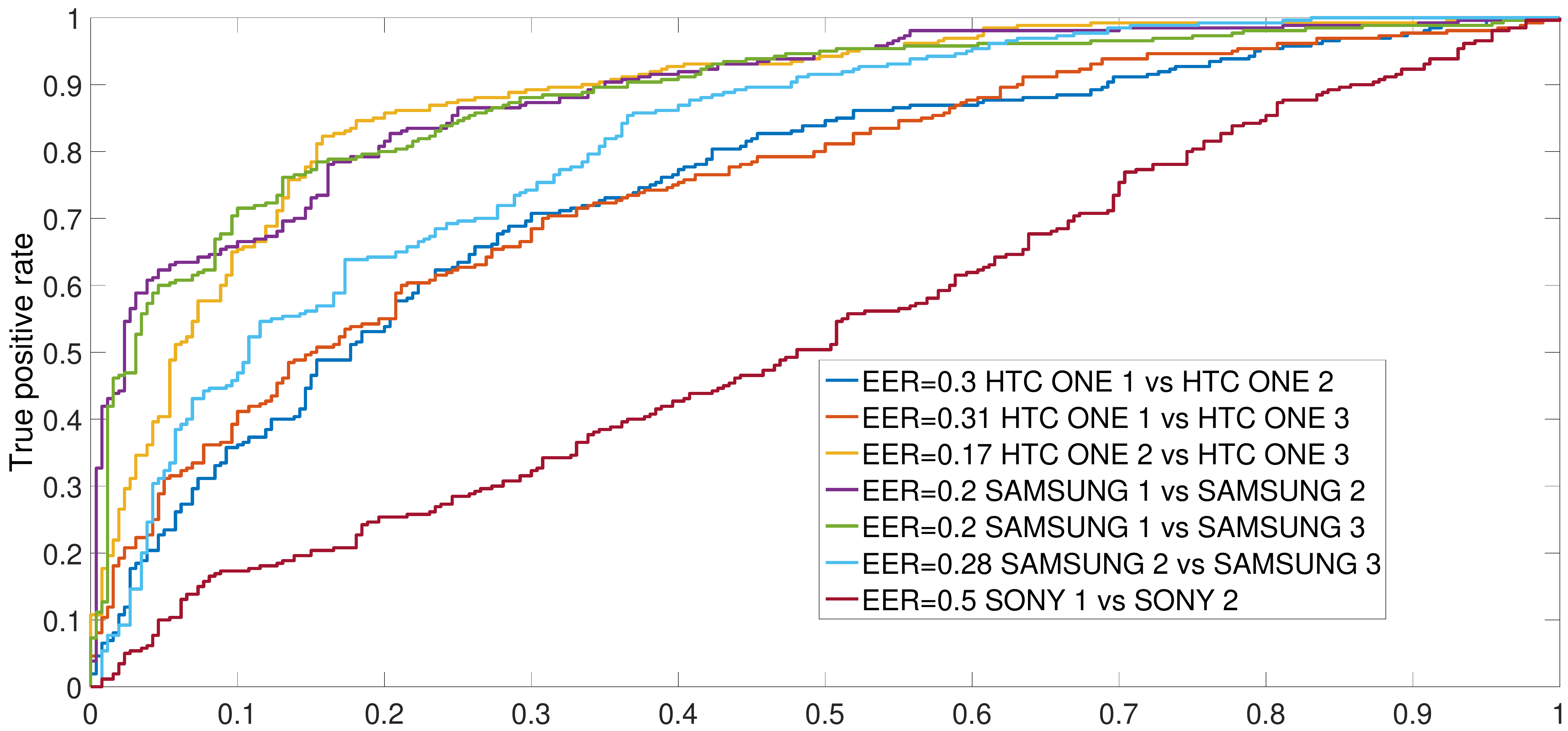}
	\caption{Intra model verification for day one}
	\label{fig:intramodelday1}
\end{figure}

Instead, figure \ref{fig:intramodelday1} shows the \ac{ROC} curves and the related \ac{EER} values for verification between mobile phones of the same model. In our case, this set includes only the HTC One (three phones), SAMSUNG (three phones) and SONY (two phones) models.
We noticed that the algorithm is able to distinguish the mobile phones only with great difficulty (EER around 0.3 and 0.2 ) for the HTC One and SAMSUNG phones, while we get random-choice between the SONY phones.

\subsection{Results based on different days of measurements}
To ensure that the results obtained in the previous section are consistent in time, we repeated the stimulation and collection of data in different days. We collected data for other two days for all the mobile phones. Note that the measurement were not taken in consecutive days but in days separated by weeks to evaluate the stability of the algorithm for a long time-frame.

We measured the stability of the algorithm on the basis of two approaches: a) we evaluated how different are the ROCs of the verification of a mobile phone against the data sets taken in three different days for another mobile phone and b) we evaluated how similar are the data sets of the same mobile phone across the three different days.
In the case a) the \ac{ROC} curves and the related \ac{EER} values should be quite similar, while in the case b) the \ac{EER} calculated from the \ac{ROC}s comparing the different data sets of the same phone should be almost random choice (EER=0.5) because they are observables taken from the same mobile phone.

\begin{figure} [!ht]
	\centering
		\includegraphics[width=\linewidth]{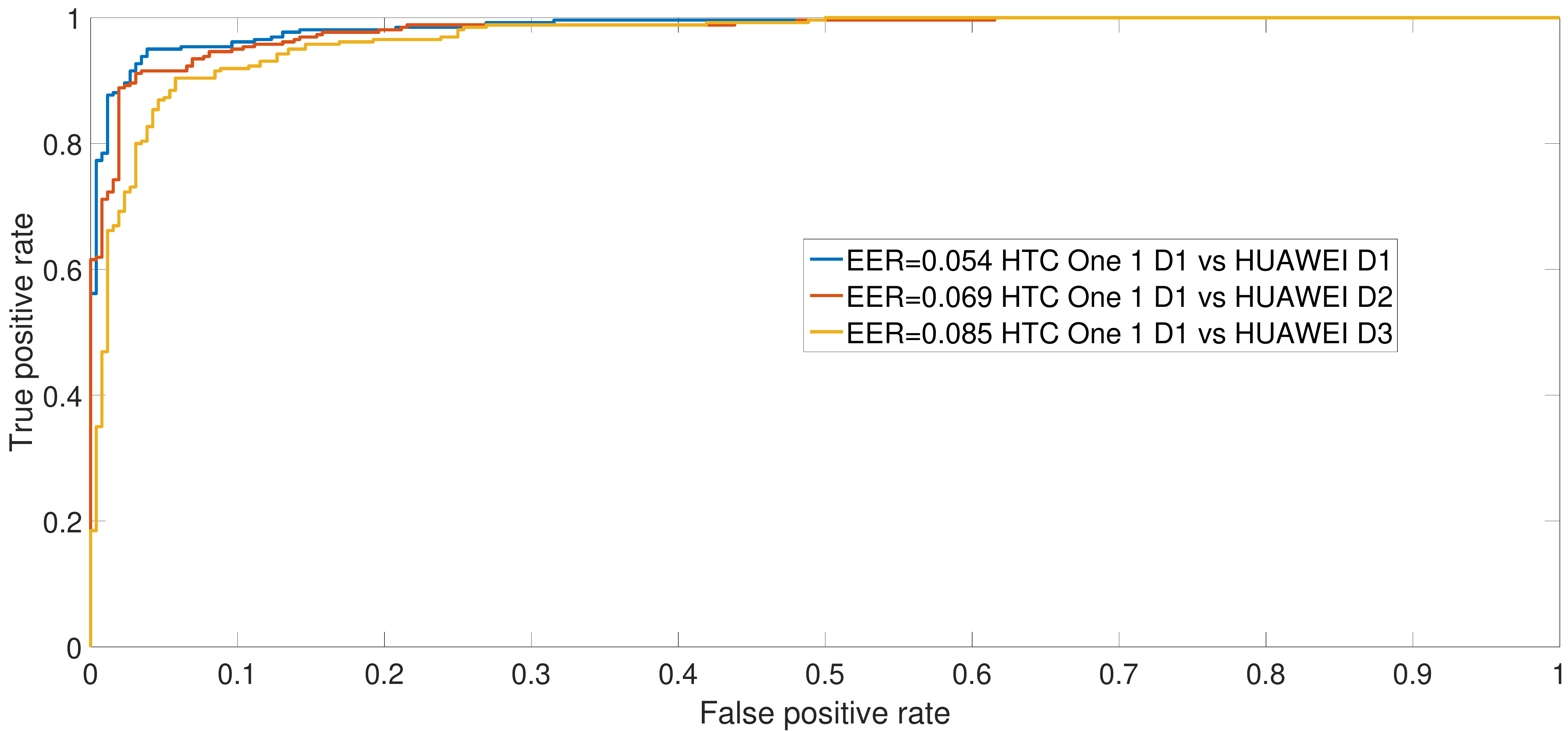}
	\caption{ROCs of HTC One against Huawei in different days}
	\label{fig:rocdiffdays}
\end{figure}

The results are shown in figure \ref{fig:rocdiffdays} for the \ac{ROC}s between one HTC One and the Huawei phone. We note that the \ac{ROC} curves are quite near. Very similar results have been obtained for the other mobile phones.

The table \ref{diffdayssamephone} shows the average \ac{EER} for all the models used in our experiments. The three values of the \ac{EER} among the combinations of days (day one against day two, day two against day three and day one against day three) have been averaged for each phone. In addition, the average among all the phones of the same model (apart from the Huawei where there is only one phone) was calculated. As expected, the resulting values of the \ac{EER} are near random-choice (EER=0.5), which shows that the stability of the approach in time.

\begin{table}
	\centering
		\begin{tabular} {l|r}
		\textbf{Mobile phone model} & \textbf{Average EER} \\ \hline
		HTC One & 0.47 \\ 
		Huawei Ascend Mate &  0.44 \\ 
		Samsung Galaxy S5 & 0.5 \\
		Sony Experia & 0.507 \\ \hline
		\end{tabular}
	\caption{Averaged EER among data from same phones in different days}
	\label{diffdayssamephone}
\end{table}

\subsection{Results in presence of Noise}
\begin{figure} [!ht]
	\centering
		\includegraphics[width=\linewidth]{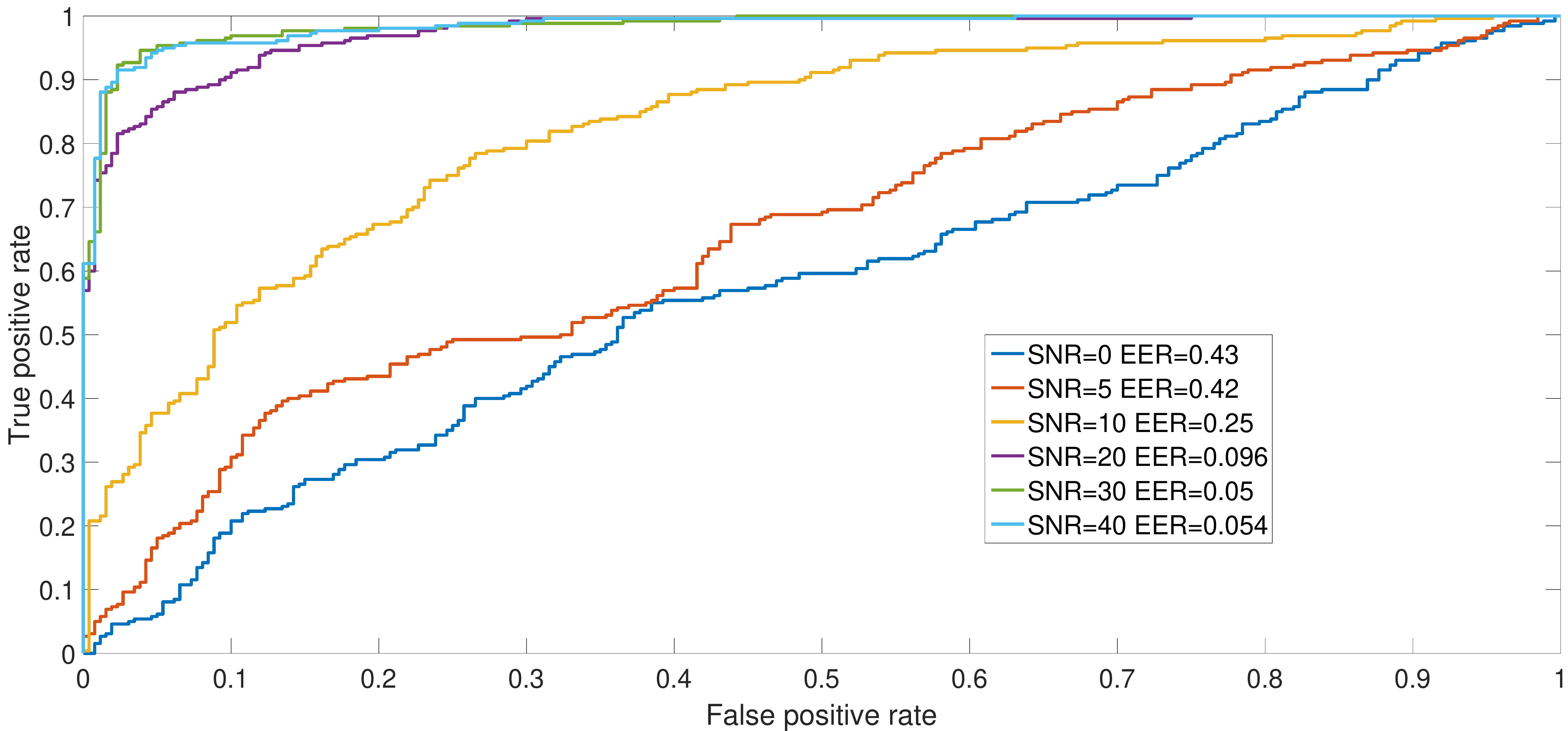}
	\caption{Verification accuracy for different SNRs of AWGN between HTC One and Huawei}
	\label{fig:rocnoise}
\end{figure}

In this section, we evaluate the impact of Gaussian Noise on the verification accuracy. \ac{AWGN} was added to the digital output collected by the magnetometers to simulate disturbances or attenuation in the propagation path between the solenoid and the mobile phone. This may be possible in a realistic application of magnetometers fingerprinting, where the position of the phone and the solenoid may not be ideal. The goal is to evaluate the loss of verification accuracy in presence of attenuation. In the simulation, \ac{AWGN} with different values of \ac{SNR} was added to the observables already collected. The process of adding \ac{AWGN} was repeated 50 times and the results were averaged to ensure the randomness of the simulation. 
The results are shown in figure \ref{fig:rocnoise}. As expected, the EER increases with the decrease of the SNR. For a value of SNR equal to 0, we obtain almost random-choice (EER=0.43).

\subsection{Results for different stimulating patterns}
\begin{figure} [!ht]
	\centering
		\includegraphics[width=\linewidth]{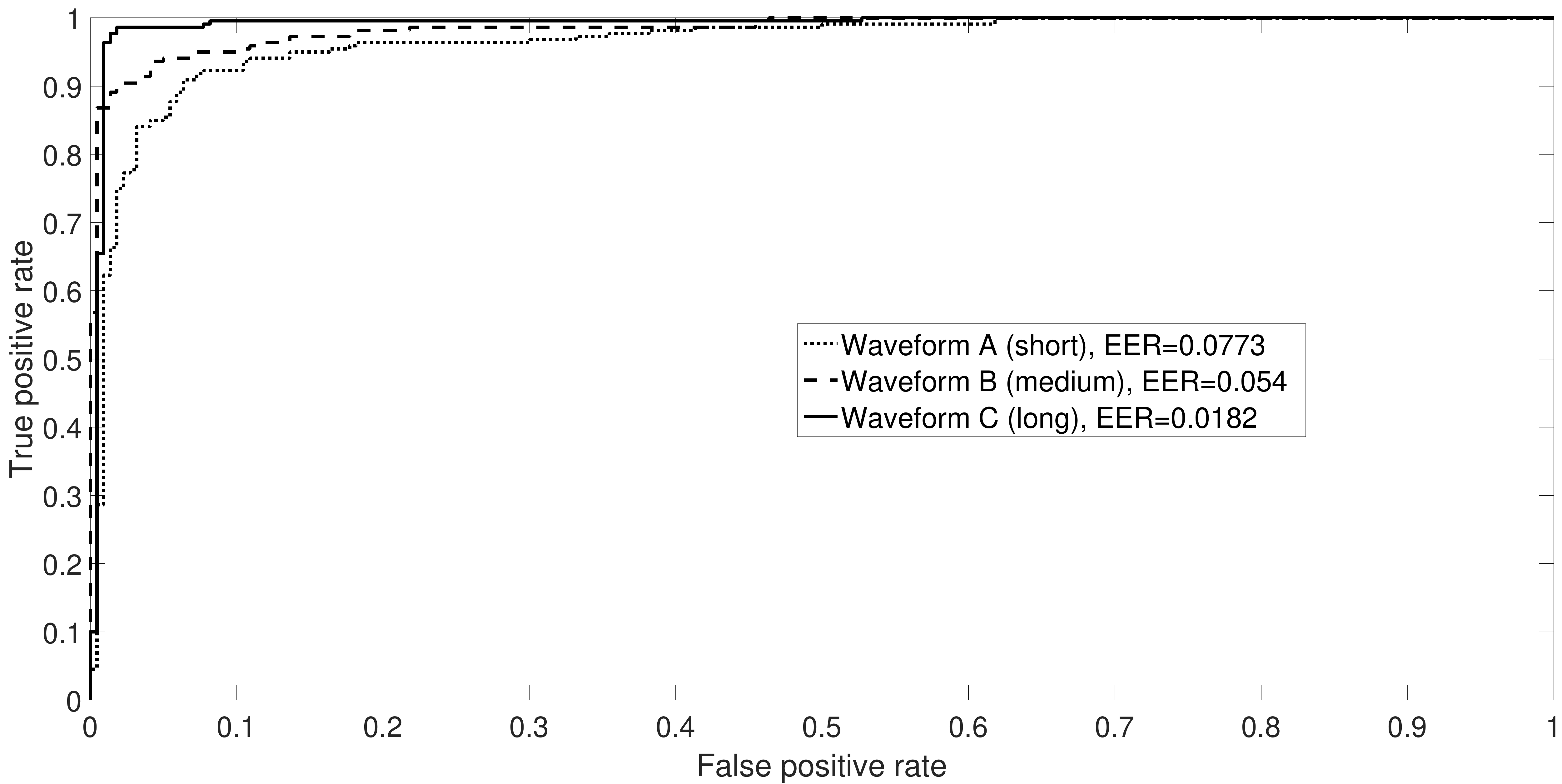}
	\caption{ROCs between HTC One 1 and Huawei for different waveforms}
	\label{fig:diffwavinter}
\end{figure}

\begin{figure} [!ht]
	\centering
		\includegraphics[width=\linewidth]{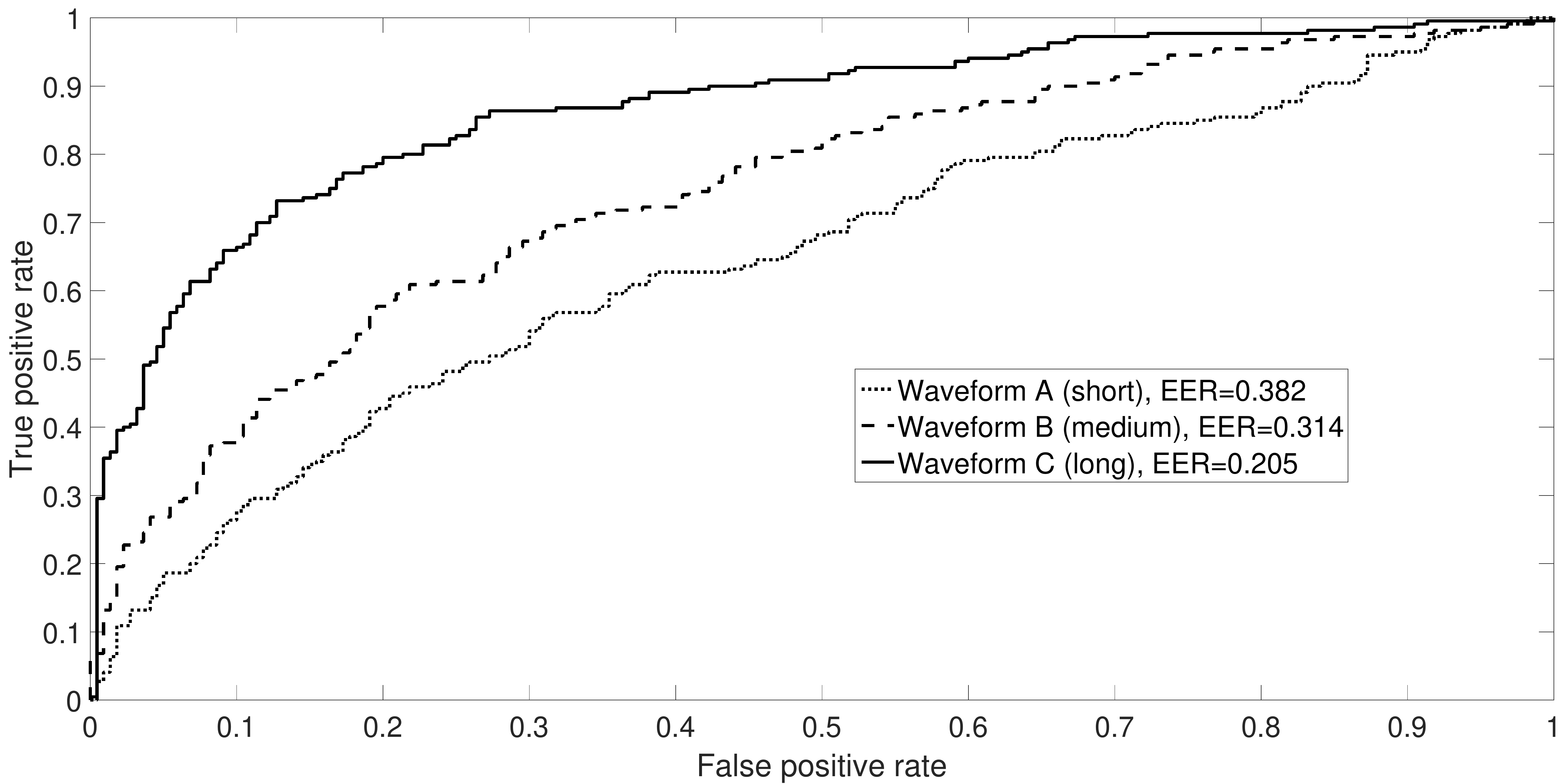}
	\caption{ROCs between HTC One 1 and HTC One 3 for different waveforms}
	\label{fig:diffwavintra}
\end{figure}

Finally, we evaluate the change in verification accuracy for different stimulating patterns: the waveforms from A to C described in figure \ref{fig:waveforms}. The goal is to evaluate if longer or more complex waveforms provide better results than shorter waveforms or vice-versa. We note that there is a trade-off because longer waveforms require longer times for the creation of the training set or for the evaluation process. In addition, the hysteresis cycle of the magnetometers require that there is a minimal time separation between the stimuli (i.e., the pulses in the waveforms of figure \ref{fig:waveforms}).
The ROCs and the average values of EERs have been calculated for the different waveforms and the results are provided in figure \ref{fig:diffwavinter} and \ref{fig:diffwavintra} for the ROCs and in table \ref{tab:eeraveragewave} for EERs.

\begin{table}
	\centering
		\begin{tabular} {l|r}
		\textbf{Intermodel} & \textbf{Average EER} \\ \hline
		Waveform A & 0.0159 \\ 
		Waveform B &  0.0136 \\ 
		Waveform C & 0.00303 \\ \hline \hline
		\textbf{Intramodel} & \textbf{Average EER} \\ \hline
		Waveform A & 0.296 \\ 
		Waveform B &  0.334 \\ 
		Waveform C & 0.195 \\ \hline \hline
		\end{tabular}
	\caption{Averaged EER for intermodel and intramodel classification}
	\label{tab:eeraveragewave}
\end{table}
 
The ROCs figures (\ref{fig:diffwavinter} and \ref{fig:diffwavintra}) show the verification between the mobile phone HTC One 1 and HTC One 3 for the intra-model case and between HTC One 1 and the Huawei mobile phone for the inter-model case. We notice that the waveform C has a slightly better performance in both cases. This result is confirmed by averaging the EERs values for all the combination of mobile phones in the inter-model and intra-model cases as shown in table \ref{tab:eeraveragewave} where Waveform C has a clear advantage over the other two waveforms, which have similar accuracy scores. 

From these results, it is clear that a longer waveform can provide a better verification accuracy at the expense of a longer training and evaluation time. Indeed, the intra-model scores for waveform C show that it would be possible to distinguish with a good level of accuracy even mobile phones of the same model and brand (i.e., intra-model).

\section{Conclusions and Future Developments}
\label{conclusions}
 
In this paper we have investigated the identification of mobile phones through their built-in magnetometers on a set of nine mobile phones, when they are stimulated by a solenoid with specific waveforms. We have shown that inter-model verification can be achieved with very high accuracy, while intra-model verification can be achieved with limited accuracy . We have evaluated and shown the impact of Gaussian Noise. We have also proven the stability of the verification results across different days. Regarding the application of different waveforms, the results show that accuracy can be improved with longer and more complex waveforms.
Future developments will extend the analysis shown in this paper with more complex waveforms and with the application of different sets of features with the goal to obtain a better intra-model accuracy.
\\
\\

\textbf{Disclosure Policy}

The authors declare that there is no conflict of interest regarding the publication of this paper.

\ifCLASSOPTIONcaptionsoff
  \newpage
\fi
\bibliographystyle{IEEEtran}
\bibliography{surveycamera}

\end{document}